\documentclass[useAMS,usenatbib]{mn2e}
\pdfoutput=1

\usepackage{amsmath}
\usepackage{amssymb}
\usepackage{threeparttable}
\usepackage{subfigure}
\usepackage{cases}
\usepackage{color}
\usepackage{graphicx}
\usepackage{hyperref}
\usepackage[figure]{hypcap}
\usepackage{tikz}

\hypersetup{colorlinks, citecolor=blue, pdfduplex=DuplexFlipLongEdge}
\hypersetup{pdftitle=HERO paper, pdfauthor=Yucong Zhu, pdfsubject=Astrophysics}

\title[HERO: A 3D General Relativistic Radiative Code]
  {HERO - A 3D General Relativistic Radiative Postprocessor for Accretion Discs around Black Holes}
\author[Y. Zhu, R. Narayan., S\c{a}dowski A., \& Psaltis D.]
  {Yucong Zhu$^1$\thanks{E-mail:\newline \hbox{yzhu@cfa.harvard.edu (YZ);} \hbox{rnarayan@cfa.harvard.edu~(RN);} \hbox{asadowsk@mit.edu~(AS);} \hbox{dpsaltis@email.arizona.edu~(DP);}}
 Ramesh Narayan$^1$\footnotemark[1],
 Aleksander Sadowski$^2$\footnotemark[1],
 Dimitrios Psaltis$^3$\footnotemark[1],
\\
  $^1$Harvard-Smithsonian Center for Astrophysics, 60 Garden Street, Cambridge, MA 02138, USA\\
  $^3$MIT Kavli Institute for Astrophysics and Space Research, 77 Massachusetts Ave, Cambridge, MA 02139, USA\\
  $^3$University of Arizona, 933 N. Cherry Ave, Tucson, AZ 85721, USA}

\begin{document}

\maketitle


\begin{abstract}
HERO (Hybrid Evaluator for Radiative Objects) is a 3D general relativistic radiative transfer code which has been tailored to the problem of analyzing radiation from simulations of relativistic accretion discs around black holes.  HERO is designed to be used as a postprocessor. Given some fixed fluid structure for the disc (i.e. density and velocity as a function of position from a hydrodynamics or magnetohydrodynamics simulation), the code obtains a self-consistent solution for the radiation field and for the gas temperatures using the condition of radiative equilibrium.  The novel aspect of HERO is that it combines two techniques: 1) a short characteristics (SC) solver that quickly converges to a self consistent disc temperature and radiation field, with 2) a long characteristics (LC) solver that provides a more accurate solution for the radiation near the photosphere and in the optically thin regions.  By combining these two techniques, we gain both the computational speed of SC and the high accuracy of LC. We present tests of HERO on a range of 1D, 2D and 3D problems in flat space and show that the results agree well with both analytical and benchmark solutions.  We also test the ability of the code to handle relativistic problems in curved space. Finally, we discuss the important topic of ray-defects, a major limitation of the SC method, and describe our strategy for minimizing the induced error.

\end{abstract}


\section{Introduction}\label{sec:intro}

Radiative transport (RT) plays a crucial role in astrophysics.
It governs our primary source of information about the cosmos, viz.,
the luminosities and spectra of cosmic sources\footnote{We do receive
  additional information from cosmic rays, neutrinos and, hopefully
  soon, gravitational waves, but these pale next to the enormous
  volume of information we receive via electromagnetic radiation.}.
In addition, radiation acts as a channel for energy transport, shaping the dynamics and
impacting the evolution of astrophysical systems on all scales: stellar
evolution, stellar and galactic winds, super-Eddington accretion,
epoch of reionization, etc.

Radiation is quite challenging to
model since it behaves in highly nonlinear and nonlocal ways.  The
radiation field is determined by a six-dimensional (6D) system of
coupled equations which link spatial positions (3D), ray directions
(2D) and frequencies (1D), all as a function of a seventh coordinate, time (in the
general time-dependent problem).  The high dimensionality of the solution vector,
combined with the complex integro-differential nonlocal nature of the
problem, results in an extremely taxing computational challenge (both in terms of
memory and computational speed).

Because of the intrinsic complexity of the problem, the earliest
radiative solvers made use of fairly restrictive approximations.  For
example, enforcing spatial symmetries allows one to greatly reduce the
dimensionality of the problem and led to the first generation of 1D/2D
radiative codes (e.g. Feautrier method that exploits symmetric and antisymmetric radiation moments, see \citealt{mihalas78}).  Several of the multidimensional RT ideas have also found application in the field of neutrino transport in core collapse supernovae \citep{burrows00,rampp02,liebendorfer04,livne04,hanke13}.  In this case, the infalling matter becomes optically thick to the neutrino flux, which propagates analagously to radiative transport in optically thick media.

3D radiative problems are extremely computationally taxing, which
motivates the need for highly efficient techniques.  Recent applications that make use of 3D RT include
models of young stellar objects \citep{wolf98}, protostellar to
protoplanetary discs \citep{indebetouw06,niccolini06}, reflection
nebulae \citep{witt96}, molecular clouds \citep{steinacker05,
  pelkonen09}, spiral galaxies \citep{bianchi08, schechtman12},
interacting and starburst galaxies \citep{chakrabarti07,hayward11},
AGNs \citep{schartmann08,stalevski12}, and cosmological reionization simulations \citep{ferland98, abel02, finlator09}. These applications are primarily concerned with point sources illuminating optically thick media.   
 
In more complex optically thick flows where extended diffuse emission
is important, the typical approach is to simplify the directional
structure of the radiation field.  One idea is to decompose the
radiation field into its moments (the first three moments being the
radiation energy density, radiation flux, and radiation pressure) and
to evolve the radiation field locally as a fluid using some
closure relation on the moments.  This approach is popular due to its
local nature and fast speed, and has been implemented in many
hydrodynamic codes \citep{turner01, bruenn06, hayes06, gonzales07,
  krumholz07, gittings08, ohsuga09, swesty09, commercon11, zhang11, kolb13}.
Flux limited diffusion (FLD, \citealt{levermore81}), which is based
on the first two moments, is the simplest
and perhaps most popular moment-based radiative method.  M1 closure is a generalization of the FLD
method which uses the first three moments. It has recently gained
traction as a fast solver for diffuse emission
\citep{levermore84,dubroca99,stone92, gonzales07, sadowski13}.
 
The field of protoplanetary disc dust modelling is one area that has
been active in developing RT techniques.  In this domain, the
monte-carlo approach has been the dominant paradigm owing to its ease
of implementation and ability to handle anisotropic scattering kernels
\citep{lopez95,niccolini03,wolf99, bjorkman01,
  pinte06}.  Monte-carlo RT has also been applied to the problem accretion discs, specifically to handle the problem of modeling Compton scattering in disc coronas \citep{dolence09,schnittman13,ghosh13} as well as dust scattering of emission from the Galactic centre \citep{odaka11}.  The primary downside of this
technique is the computational expense and noise associated with
photon statistics. The method is also limited to modest optical depths. 

The limitations of the various methods mentioned above led to the
development of more deterministic discretized finite-differencing
methods \citep{stenholm91, steinacker02} and discrete ordinates
characteristic methods \citep{dullemond00, steinacker02, hayes03, vogler05, heinemann06, woitke09, hayek10,
  davis12, jiang14}, which have a number of widely discussed advantages. However, all
of these methods suffer from the phenomenon of "ray-defects" (see
appendix \ref{app:raydefects} for a detailed discussion).  Another
twist to the discretization strategy involves expanding and evolving
the radiation field as a combination of spherical harmonics
\citep{szucheng82, evans97, mcclarren08}.  Spherical harmonics respect
rotational symmetry and therefore do not suffer from the linear ray defect
patterns that plague other discrete ordinate methods.  The tradeoff is
that spherical methods typically suffer from artificial sidelobe
patterns due to the finite order cutoff used in taking the series
expansion of the radiation field.
 
Raytracing methods are also used for predicting the observed flux from astrophysical objects. These codes typically do not solve for the global radiation field within an object; instead the focus is on calculating the apparent intensities that reach a distant observer.  In these codes, the typical assumption is to ignore scattering (i.e. to ignore nonlocal coupling of the radiation field) since this allows one to immediately compute the evolution of ray intensity by simply integrating the local emissivities along a photon geodesic (e.g. \citealt{cunningham73, ozel01, huang07, dexter09, shcherbakov11, vincent11, chan13, bohn14}).

Despite the multitude of radiative solvers available, none are currently
able to solve the problem of optically thick emission from accretion
discs around black holes (BHs).  The main difficulty here is that a general relativistic 3D framework is needed to
properly account for both light bending effects and
doppler/gravitational redshifts.  The work reported here is a first attempt at
tackling the radiation problem in full glory around black holes.  The
code we describe here called HERO is a \emph{postprocessor} --  given some fixed gas structure (density, velocity,
energy injection rate), as determined by a separate BH accretion
disc simulation, our code \emph{solves for the radiation field along with its self-consistent temperature
solution}. We explain here how our code works and show verification
tests to gauge its performance under various conditions.  

The organization of the paper is as follows. In \S\ref{sec:RT}, we first describe the
method by which HERO operates.  We give a brief overview of the
radiative transfer problem that HERO solves, and how it generalizes
in curved space.  We also detail how the two radiative solvers
(short/long characteristics) operate, and the numerical methods
involved (i.e. acceleration schemes, interpolation, discretization strategy).
Next, in \S\ref{sec:results} we check the code by comparing to
analytic and benchmark results for 1D, 2D, and 3D test configurations.
Finally, in appendix \S\ref{app:raydefects} we end with a discussion
of ray-defects, a systematic problem that plagues short-characteristic
radiative solvers.



\section{Radiative Solver}\label{sec:RT}


Our radiative code consists of three primary components: 

\begin{enumerate}
\item A \textbf{short characteristic solver} for quickly obtaining an approximate solution to the radiation field
\item A \textbf{long characteristic solver} for more accurate modeling of the radiation field
\item An optional \textbf{raytracer} for generating mock observations from some distant observing plane
\end{enumerate}



We employ a {\it hybrid} approach to solve the radiation field, hence
the name of the code: HERO (Hybrid Evaluator for Radiative
Objects).  The code is hybrid in the sense that it uses both
(i)``short characteristics" (SC, \citealt{mihalas78}) and (ii)``long
characteristics" (LC, \citealt{feautrier64}).  Here ``short'' and ``long'' refer to the length of the light rays that are
considered in each iteration of the solver.  The short characteristics
method is only concerned with propagation of radiation from a
given cell to its immediate neighbours, whereas the long characteristic
method traces rays all the way to the edge of the computational grid.

The motivation for developing a hybrid approach is to allow for the accurate
modelling of radiation in optically thin regions via long
characteristics, while retaining the computational speed offered by
short characteristics in optically thick regions (see
\S\ref{sec:shortChar} \& \S\ref{sec:longChar} for a detailed
comparison of the methods).  For any given problem, we first apply
short characteristics to quickly solve for the local
radiation/temperature. We then feed this solution as the initial guess
for a more detailed long characteristics calculation.  Using such a
hybrid approach allows us to combine the ``best of both worlds." 

In addition to solving for the radiation field, in both SC and LC, we also solve for a
self-consistent gas temperature.  The
general idea is to loop back and forth between solving for the
radiation field given a fixed temperature structure, and solving for
the equilibrium temperature distribution given a fixed radiation
field. This procedure is iterated until convergence, leaving us with
a self-consistent solution for both the radiation and the
temperature.

Our goal in developing HERO is to model/investigate the observed
properties of relativistic accretion discs around black holes, and to compare the radiative properties of simulated discs with data collected by earthbound telescopes.  Therefore, after completing the SC and LC
steps described above, the final step is to input the radiative and
temperature structure as computed from LC and to generate synthetic
observations of the disc as seen by a distant observer.  For this
stage we use the standard raytracing approach (see \S\ref{sec:raytracing} for details).

Due to the high temperatures present in accretion discs, Comptonization plays a crucial role in determining the observed radiation from relativistic disc photospheres.  Because of the complex nature of the Compton scattering kernel, we defer discussion of Comptonization to a follow-up paper, which will focus exclusively on explaining and testing our relativistic scattering module.

\subsection{Short Characteristics}\label{sec:shortChar}

\begin{figure}
\begin{center}  
\includegraphics[width=0.35\textwidth]{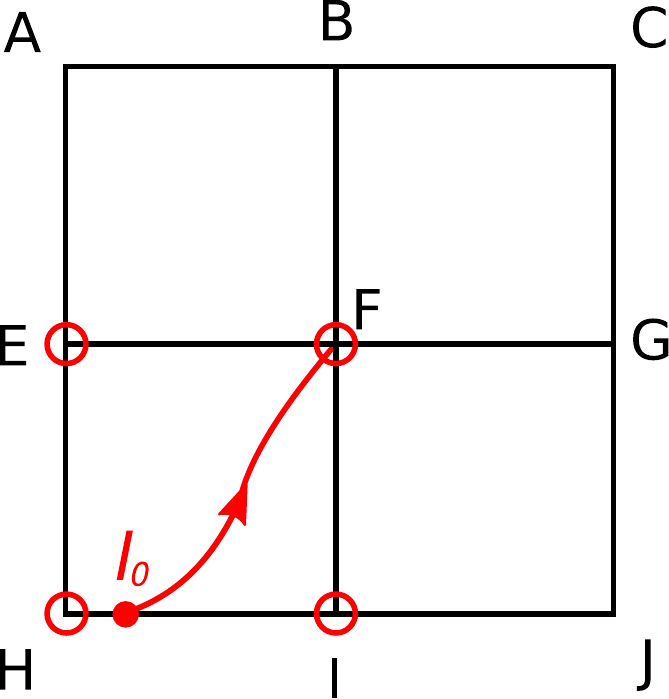}
\caption{Schematic of the short characteristics method.  A curved ray (null geodesic) is shot ``upstream'' to determine the intensity at F.  This ray is terminated at the neighbouring cell boundaries, and the intensity $I_0$ at the boundary is computed by interpolation of neighbouring grid points (in this case points H and I).  The source function $S(\tau ')$ in Eq. \ref{eq:shortCharRT} is also obtained by interpolation on neighbouring points (points E, F, H, I), and the intensity at F is thereby calculated.  The procedure is repeated for a number of rays in different directions to obtain an estimate of the radiation field at F.
\label{fig:ShortCharSchematic}}
\end{center}  
\end{figure}  

Short characteristics is a popular algorithm for tackling multidimensional radiative transfer problems \citep{mihalas78,olson87,kunasz88}. It is a local method and hence very fast. The basic idea is to solve for the radiation field at a given point (reference cell) using only the information provided by neighbouring cells (see schematic in Figure \ref{fig:ShortCharSchematic}).

\subsubsection{Ordinary Radiative Transfer Equation}\label{sec:setup}

To set the stage, we first discuss the standard case of flat space, where the radiative transfer equation takes the form:
\begin{equation}
\frac{dI_{\nu}}{ds} = -(\kappa_{\nu} + \sigma_{\nu}) I_{\nu} + j_{\nu} + \sigma_{\nu} \int \phi_{\nu}(\Omega, \Omega')I_{\nu}(\Omega')d\Omega'.\label{eq:RTstart}
\end{equation}
Here $I_\nu$ is the intensity of a ray travelling along path $s$
directed towards $\Omega$; $\kappa_\nu$ and $\sigma_\nu$ are the
absorption and scattering coefficients; $\phi_{\nu}$ is the scattering
shape function, normalized such that $\int \phi_{\nu}(\Omega,
\Omega')d\Omega' = 1$; and $j_{\nu}$ is the emission coefficient.
Typically, Eq.\ref{eq:RTstart} is simplified to the form
\begin{equation}
\frac{dI_{\nu}}{d\tau_{\nu}} = -I_{\nu} + S_{\nu},
\end{equation}
where the the optical depth $\tau_\nu$ is given by
\begin{equation}
d\tau_{\nu} = (\kappa_{\nu} + \sigma_{\nu}) ds,
\end{equation}
and the source function $S_{\nu}$ is defined as
\begin{equation}
S_{\nu} = \frac{j_{\nu} + \sigma_{\nu} \int \phi_{\nu}(\Omega, \Omega')I_{\nu}(\Omega')d\Omega'}{\kappa_{\nu} + \sigma_{\nu}}. \label{eq:SourceFunction}
\end{equation}
In the case of isotropic scattering, $\phi_\nu = 1/4\pi$, and the
scattering term simplifies to
\begin{equation}
\int \phi_{\nu}(\Omega, \Omega')I_{\nu}(\Omega')d\Omega' = \frac{1}{4\pi} \int I_{\nu}(\Omega')d\Omega' \equiv J_{\nu}, \label{eq:Jdef}
\end{equation}
where $J_\nu$ is the angle-averaged mean intensity. We assume isotropic scattering
in all of the work reported here.
We further assume a thermal source, for which the
emissivity is given by $j_\nu = \kappa_\nu B_\nu(T)$, where $B_\nu(T)$
is the Planck function corresponding to the local temperature $T$ of
the medium.  With these two simplifications, the source function can be
compactly rewritten as:
\begin{equation}
S_{\nu} = \epsilon_\nu B_\nu + (1-\epsilon_\nu) J_\nu, \label{eq:simpleSource}
\end{equation}
where $\epsilon_\nu$ is the photon interaction destruction probability, given by the ratio of the absorption and total opacity:
\begin{equation}
\epsilon_\nu \equiv \frac{\kappa_\nu}{\kappa_\nu + \sigma_\nu}.\label{eq:epsilon}
\end{equation}
To avoid confusion with summation indices, in the remainder of this paper we will no longer use a subscript $\nu$ to note the frequency dependence of radiative quantities.  Unless otherwise stated, this will apply to all subsequent intensities, opacities, source functions, and emissivities.

\subsubsection{Covariant Formulation of the Radiative Transfer Equation}\label{sec:setup}

Because HERO has been developed for relativistic problems, we use a generalization of the radiative transfer equation that accounts for relativistic effects such as light bending and doppler boosting.  The generalization of the radiative transfer equation to curved space has been discussed in the literature \citep{mihalas84,lindquist66}, but we give a brief primer here for completeness. Starting from the time dependent version of Eq. \ref{eq:RTstart}:
\begin{equation}
\frac{1}{c}\frac{dI}{dt} + (\hat{n}\cdot\nabla)I = G, \label{eq:stdRT}
\end{equation}
where $G$ on the righthand side contains all scattering and absorption source terms that interact with the radiation field, and $\hat{n}$ is a normalized 3-vector denoting the propagation direction of photons.  We can rewrite Eq. \ref{eq:stdRT} in a more revealing form by introducing the direction 4-vector $k^\alpha$ whose orthonormal form is (note that $k\cdot k=0$):
\begin{equation}
k^\alpha = (1,\hat{n}).
\end{equation}
This lets us rewrite the RT equation (setting $c=1$) as:
\begin{equation}
k^\alpha \frac{d}{dx^\alpha} I = G \equiv -(\kappa + \sigma) I + j + \sigma J .\label{eq:startCovRT}
\end{equation}

By considering the Lorentz transformation properties of the various quantities, Eq.\ref{eq:startCovRT} can be recast in an invariant form (see \citealt{mihalas84}, \S7.1):
\begin{equation}
p^\alpha \frac{\partial}{\partial x^\alpha} \mathcal{I} = \mathcal{G}, \label{eq:specialRT}
\end{equation}
where the photon 4-momentum $p^\alpha = h \nu k^\alpha$ takes on
the role of the propagation 4-vector. Instead of the usual definitions of intensity $I$ and source term $G$, this equation considers
their corresponding relativistically invariant versions, 
\begin{equation}
\mathcal{I} = I/\nu^3, \qquad \mathcal{G} = hG/\nu^2,
\end{equation}
where the scaling of $I$ is determined by photon number conservation in phase space. The $(h\nu)^{-2}$ scaling for $G$ arises after transforming Eq.\ref{eq:startCovRT} to use $\mathcal{I}$ and allows us to construct the invariant scalings for the absorption and scattering coefficients as:
\begin{equation}
\mathfrak{k}_\nu = h \nu \kappa_\nu, \qquad
\mathfrak{s}_\nu = h \nu \sigma_\nu,
\end{equation}
and for the emissivity and source function,
\begin{equation}
\mathfrak{j}_\nu = h j_\nu/\nu^2, \qquad \mathcal{S} = S/\nu^3.
\end{equation}
In terms of these new quantities, the invariant source $\mathcal{G}$ becomes
\begin{equation}
\mathcal{G} = -(\mathfrak{k} + \mathfrak{s})(\mathcal{I} - \mathcal{S}).
\end{equation}
For convenience, since we usually work with comoving quantities, $\kappa_{\rm{c}}$, $\sigma_{\rm{c}}$, $S_{\rm{c}}$, we can instead write $\mathcal{G}$ as:
\begin{align}
\mathcal{G} &= -h\nu_{\rm{c}} (\kappa_{\rm{c}} + \sigma_{\rm{c}}) \left[-\mathcal{I} + \frac{S_{\rm{c}}}{\nu^3_{\rm{c}}}\right] \nonumber\\ 
& = p^\alpha u_\alpha  (\kappa_{\rm{c}} + \sigma_{\rm{c}}) \left[ -\mathcal{I} + \frac{S_{\rm{c}}}{\nu^3_{\rm{c}}}\right] , \label{eq:comovingG}
\end{align}
where $u^\mu$ is the four-velocity of the fluid.  Eq. \ref{eq:specialRT} is in manifestly covariant form and the generalization to curved space is simply accounted for by replacing the directional derivative $\partial/\partial x^\alpha$ with the covariant derivative:
\begin{equation}
\frac{D}{Dx^\alpha} = \frac{\partial}{\partial x^\alpha} - \Gamma_{\alpha \beta}^{\gamma} p^\beta \frac{\partial}{\partial p^\gamma} ,\label{eq:covariantDerivative}
\end{equation}
evaluated along a null geodesic \citep{lindquist66}.  The general covariant form of the RT equation then takes the form:
\begin{equation}
p^\alpha \frac{D}{Dx^\alpha} \mathcal{I} = \mathcal{G} \label{eq:covRT}
\end{equation}
We can further simplify the above expression by using the photon geodesic equation,
\begin{equation}
\frac{dp^\gamma}{d\lambda} = -\Gamma_{\alpha \beta}^{\gamma} p^\alpha p^\beta, \label{eq:geodes}
\end{equation}
where $\lambda$ is an affine parameter defined by $p^\alpha = dx^\alpha/d\lambda$.  Using Eq. \ref{eq:geodes} to eliminate the Christoffel symbols in Eq. \ref{eq:covariantDerivative}, the RT equation becomes:
\begin{equation}
p^\alpha \frac{\partial \mathcal{I}}{\partial x^\alpha} + \frac{dp^\alpha}{d\lambda} \frac{\partial \mathcal{I}}{\partial p^\alpha} = \mathcal{G}. 
\end{equation}
Finally, using the chain rule for partial differentiation, we obtain the very simple equation
\begin{equation}
\frac{d \mathcal{I}}{d\lambda} = \mathcal{G}. \label{eq:covRTsimple}
\end{equation}

This is the form of the covariant RT equation that we employ in HERO.  The only point that requires some care is choosing the correct frequency when evaluating the source term $\mathcal{G}$.  For instance, if we wish to evaluate $\mathcal{I}$ at frequency $\nu$ in the lab frame, but we compute $\mathcal{G}$ in the fluid comoving frame (the natural
frame since the opacity is defined there), then $\nu_{\rm{c}}$ in Eq. \ref{eq:comovingG} must be chosen such that it corresponds to the lab frame $\nu$ used to label $\mathcal{I}$.

\subsection{Implementation of Short Characteristics}

Given a ray in the reference cell F, we compute the upstream location where it intersects the boundary of that cell (see Fig.~\ref{fig:ShortCharSchematic}) and use this location to determine the relative weights contributed by each of the four nearest neighbours on that boundary. In flat space, it is possible to write down an analytical formula for the intersection point (e.g. \citealt{dullemond00}), but there is no generalization in terms of simple functions for curved space.  HERO does the calculation numerically. 

Although this is computationally somewhat expensive, it needs to be done only once as part of the initialization steps. The information from this initial computation is saved and used repeatedly during the SC iterations.  Given the intersection of a ray with the nearest neighbour cell boundary, we compute the intensity of this ray in the reference cell F using the radiative transfer equation (Eq. \ref{eq:covRTsimple}):
\begin{equation}
I_{F} = I_{0}\exp(-\tau_0) + \int_0^{\tau_0} S(\tau') \exp(-\tau') d\tau' \label{eq:shortCharRT},
\end{equation}
where $I_0$ is the incoming intensity at the boundary, and $\tau_0$ is the total optical depth of the photon path leading from the boundary to point F (computed using the average of the central and boundary opacities).  The intensity $I_{0}$ is computed by interpolating from the four boundary points.  Our current implementation of short characteristics uses linear interpolations at various stages (this is easily improved in the future).  In this spirit, the integral evaluation in Eq. \ref{eq:shortCharRT} is computed as
\begin{align}
&\int_0^{\tau} S(\tau') \exp(-\tau') d\tau' \nonumber \\
&= \frac{e^{-\tau}(S_0 - S_\tau(1+\tau)) + S_0(\tau - 1) + S_\tau}{\tau} , \label{eq:linearSource}
\end{align}
which is the analytic solution when the source function $S$ varies linearly with $\tau$ along the photon trajectory.  Higher order interpolation schemes have been explored in other codes (e.g. \citealt{kunasz88}); however, special care is needed to ensure non-negativity of the source function.

Our implementation of SC in HERO retains some memory of the radiative quantities from the past iteration.  This helps with stability.  Thus, in computing the intensity at iteration $n$, we use the linear combination 
\begin{equation}
I^{n} = (1-m) I^{n-1}_{FS} + m I^{n-1}. \label{eq:averagingIteration}
\end{equation}
where $I^{n-1}_{FS}$ is the result from the RT formal solution (i.e. evolving the radiation field through SC). Typically, we set  $m=0.5$.  HERO with SC converges within a reasonable number of iterations $\sim10-1000s$, depending on the degree of scattering/coupling in the problem.

The above discussion assumed that the gas temperatures are given. More often, one does not know the temperature but has to solve for it based on boundary conditions and internal heat sources. In this case, between every two of the above iterations for the intensity, the code carries out a round of iterations (typically $\sim10$s) to solve for the temperature using the condition of radiative equilibrium, i.e., the requirement that the heating and cooling rates of the gas should balance.  Radiative equilibrium requires at each position $x$
\begin{equation}
Q_{\rm cool}(x) = Q_{\rm heat}(x) + Q_{\rm inject}(x),
\end{equation}
where
\begin{align} \label{eq:Tequilibrium}
Q_{\rm{cool}}(x) &= \int  \kappa_\nu(x) B_\nu[T(x)] d\nu\ ,\ \nonumber \\
\quad Q_{\rm{heat}}(x) &= \int  \kappa_\nu(x) J_\nu(x) d\nu,
\end{align}
and $Q_{\rm{inject}}(x)$ refers to any additional injection of energy into the fluid, e.g., through viscous dissipation
in an accretion disc.

\subsection{Long Characteristics}\label{sec:longChar}

\begin{figure}
\begin{center}  
\includegraphics[width=0.4\textwidth]{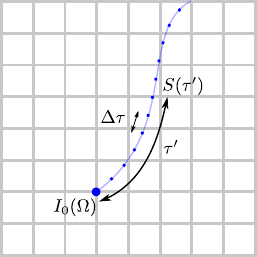}
\caption{Schematic of long characteristics RT solver.  We calculate the local mean intensity at an observing cell (large blue circle) by integrating the source function along photon geodesics (light blue path).  The RT integral is evaluated by direct summation of contributions from linear piece-wise segments (small blue dots).
\label{fig:longSchematic}}
\end{center}  
\end{figure}  

Similar to the short characteristics method, our long characteristics solver obtains the radiation field by evolving ray intensities according to the radiative transfer equation (c.f. Eq. \ref{eq:covRTsimple}).  Figure \ref{fig:longSchematic} shows a schematic of our approach to LC -- rays are shot upstream, and we evaluate the intensity at the observing cell according to
\begin{equation}
I_{0} = I_{b}\exp(-\tau_b) + \int_0^{\tau_b} S(\tau') \exp(-\tau') d\tau'. \label{eq:longCharRT}
\end{equation}
The above expression is identical to our SC calculation (Eq.\ref{eq:shortCharRT}), except that $I_b$ now represents the intensity distribution at the grid boundaries and the ray integral continues past the immediate neighbouring cells to traverse the entire spatial domain.  The integral in Eq. \ref{eq:longCharRT} is evaluated in discrete steps ($\tau' \rightarrow \tau' + \Delta \tau$), where each source-function piece is evaluated using the same linear interpolation scheme as used in SC (Eq. \ref{eq:linearSource}).

One major difference between our LC and SC solvers is the choice of angular grid. In SC we used a fixed angular resolution ($N_A = 80$, see \S\ref{sec:3Dangles}) for all cells.  This fairly low resolution choice is adequate (though only barely) in SC since one only needs to resolve the  neighbouring $27$ grid cells.  In LC, since the rays traverse the entire spatial domain, much higher angular resolution is needed to resolve the emission from distant source cells.  This becomes particularly problematic with spherical coordinates where the grid spans a few decades in radius. Cells located at larger radii have a difficult time ``seeing'' the inner regions unless one takes care in choosing the ray directions.

Our solution is to dynamically set the angular resolution used by the LC solver such that there is sufficient coverage to resolve all the cells within the domain.  Consider, as a typical example, a 3D grid in spherical coordinates which extends from an inner radius $r_{\rm in}$ to an outer radius $r_{\rm out} \gg r_{\rm in}$. The radial dynamic range may be up to three decades, as in some of the examples described later. Let us suppose that our spherical spatial grid uses uniform angular spacing in $\theta$, with $N_\theta$ cells, and  logarithmic spacing in radius with $N_r \lesssim N_\theta$ cells per decade. Consider a cell at some radius $r$. For this cell, the apparent density of other cells as viewed in its local "sky" is highly anisotropic. For rays traveling from the direction of the coordinate centre, this cell needs to sample the local radiation field with an angular resolution of order $\Delta\theta_{\rm res} \sim r_{\rm in} / rN_\theta$ in order to adequately sample the contribution of all the cells at the centre. In the opposite direction, however, an angular resolution of order $ \sim1/N_\theta$ is sufficient. 

The above requirement can be achieved by designing a suitable angular tiling of the sky.  One simple prescription is to vary the angular resolution $\Delta \theta_{\rm res}$ as a function of direction $\chi$ relative to the local radius vector as follows:
\begin{equation}
\Delta \theta_{\rm res} = \frac{f}{N_\theta}  \left( \frac{r_{\rm in}}{r} + \chi \right), \label{eq:tiling}
\end{equation}
where $f$ controls the factor by which we oversample to reduce noise.  Usually $f=1$ is adequate, however even with this relatively low resolution, the number of tiles varies from $\sim N_\theta^2$ for the innermost radial grid cells to several orders of magnitude larger for outer cells.  The increase in resolution at the outer cells is needed in order to resolve the interior. The average number of rays per cell is thus a factor of 100 or more larger than the $N_A=80$ rays that we consider with short characteristics. In addition, each ray in LC traverses the entire grid, compared to just one cell with SC. For both reasons, one iteration of LC often takes $10^4$ times more computation time than one iteration of SC!  On the other hand, one iteration of LC moves the solution much closer to convergence since information is propagated across the entire grid.  

The main weakness of LC is its inability to handle optically thick regions in scattering-dominated problems. This is because of the huge number of iterations needed to model properly the diffusion process.  Since LC cannot be feasibly run for more than a few dozen iterations, it cannot be used to update radiative information deep inside optically thick objects.  Luckily, it is precisely in this optically thick regime where short characteristics is free from the ray-defect problem and can be trusted to produce unbiased results.  This motivates our hybrid scheme: first we run many iterations of short characteristics to resolve the complete diffusive process and to correctly obtain the radiation field in the optically thick parts of the problem;  second we run a few passes of LC to ensure that the optically thin regions of the radiation field are handled accurately and are devoid of ray-defects.


\subsection{Acceleration Schemes}

The radiative quantities $S$ and $J$ are inextricably linked.  For future reference, we quantify the dependence of $J$ on $S$ by means of a $\Lambda$ operator, defined such that
%
\begin{equation}
J_i = \Lambda_i^j S_j,\label{eq:lambdaDef}
\end{equation}
where the sub/superscripts denote grid cells.  Each iteration of the radiative solver\footnote{In both SC and LC, the mean intensity term $J$ is evaluated by averaging $I$ via Eq.\ref{eq:Jdef}}  acts as $\Lambda$, describing how $J$ is related to $S$.  However, since $S$ itself has a scattering term, it needs to be updated along with $J$.  Writing this explicitly (c.f. Eq. \ref{eq:simpleSource}),
\begin{align}
S^{n+1}&=(1-\epsilon) J^n + \epsilon B \nonumber\\
&=(1-\epsilon) \Lambda[S^{n}] + \epsilon B. \label{eq:LI}
\end{align}
By switching between solving for $J$ as a function of $S$ (Eq. \ref{eq:lambdaDef}),  and for $S$ as a function of $J$ (Eq. \ref{eq:LI}), we seek to find a stationary self-consistent solution for all our radiative quantities.  Unfortunately, simply applying these two transformations one after the other and iterating has well known convergence issues whenever scattering strongly dominates (say with $\epsilon < 10^{-2}$).  

To avoid slow convergence, acceleration schemes are often employed.  Accelerated lambda iteration (ALI) is one such technique.  To understand the idea behind ALI, rewrite Eq. \ref{eq:LI} so as to isolate $S$:
\begin{equation}
S=[1-(1-\epsilon)\Lambda]^{-1}[\epsilon B] \label{eq:solveS}.
\end{equation}
Formally, one can obtain the complete solution by evaluating the inverse matrix on the right hand side of Eq.\ref{eq:solveS} and calculating $S$ from $B$.  Then evaluating Eq. \ref{eq:lambdaDef} gives $J$.  However, in 3D, $\Lambda$ is an enormous matrix and it is impractical to carry out full matrix inversion.  Instead, ALI considers an \emph{approximate} lambda operator, typically chosen to be the diagonal component $\Lambda_{ii}$ (i.e. the contribution to the local $J$ from solely the local $S$ -- usually the largest component).  
Denoting the shift in our old formal solution without acceleration as $\Delta S^{FS} = S^{n+1}_i - S^{n}_i$, solving the system taking into account $\Lambda$ yields (see \citealt{hubeny03} for details):
\begin{align}
\Delta S_i^{ALI} &= \frac{\Delta S^{FS}}{1 - (1-\epsilon_i)\Lambda_{ii}} .
 \label{eq:ALI}
\end{align}
To obtain the diagonal of the $\Lambda$ operator, we consider:
\begin{equation}
\Lambda_{ii} = \frac{J_{ii}}{S_i} ,
\end{equation}
where $J_{ii}$ represents the local contribution to $J$ from the source function within the reference grid cell.  This self-illumination contribution to $J$ is the following sum over the $N_A$ ray angles considered:
\begin{equation}
J_{ii}= \frac{1}{N_A}\sum\limits_{i=1}^{N_A} I_i^{local},
\end{equation}
where $I$ is evaluated by summing up all local sources within the cell:
\begin{equation}
I_i^{local} = \int_0^{\tau_0} S(\tau') \exp(-\tau') d\tau'  .\label{eq:RTpathevaluation}
\end{equation}
ALI is crucial for scattering-dominated problems whenever $\epsilon < 0.01$ -- see convergence tests in \S\ref{sec:convergence}.
Even with ALI, we often need to run hundreds of iterations before the solution settles down and converges.  This is feasible with SC, which is quite fast, but not practical with the much slower LC. The convergence criterion that we use is:
\begin{equation}
\textrm{max}\left(\frac{|\Delta S_i|}{S_i}\right) \le \delta_c  ,\label{eq:convCriteria}
\end{equation}
where the convergence level $\delta_c$ is preset to a desired accuracy level (typically $\delta_c = 10^{-4}$).  One caveat with this criterion is that it only indicates the level at which HERO is internally satisfied with the solution (i.e. that HERO converged onto a solution).  HERO can still converge onto the wrong solution due to systematic errors caused by the numerical setup (i.e. the assumption of linearity for how $S$ varies with $\tau$, or the assumption that $I(\Omega)$ has linear dependence on $\Omega$ across neighboring cells).  We find that in practice, these systematic effects lead to the HERO solution being biased at the $\sim5\%$ level when compared with exact analytical results.

\subsection{Raytracing}\label{sec:raytracing}

At the end of the day, we wish to calculate images and compute synthetic spectra from our radiating objects.  This is accomplished by means of raytracing from some distant observation plane.  We consider a large
number of rays arriving at the observer and trace each backwards
in time to determine its trajectory and point of origin.  Then, using
the radiative transfer equation and the source function as calculated
via the LC method, we compute the observed intensity of each ray. Finally, we
combine the rays to construct the observed image of the disc
as well as the spectrum of the source. These synthetic results, which are easily calculated for different viewing angles, can be directly compared to observations of accretion discs.

Photon trajectories are handled by numerically evaluating the null geodesic equation Eq. \ref{eq:geodes} (a second order PDE in $x$ and $p$): 
\begin{align}
\frac{d^2 x^\alpha}{d\lambda^2} &= - \Gamma^\alpha_{\beta \gamma} p^\beta p^\gamma \nonumber\\
\text{with} \; p^\alpha &= \frac{dx^\alpha}{d\lambda}.
\end{align}
The emission seen at the observer plane is obtained by summing up the emissivity along the ray paths (same exercise as Eq. \ref{eq:longCharRT}).  This yields a grid of ray intensities ("image" of the system) as projected on the observation plane.  Integrating all the rays generates the final observed spectrum via
\begin{equation}
F_\nu = \int I_\nu cos\theta d\Omega,
\end{equation}
where $\theta \approx 0$ is the normal incident angle of the ray at the observer plane, $I_\nu$ is the intensity at the observer plane, and $d\Omega$ is the angular size of a single pixel at the observer plane.

\subsection{Frequency Discretization}

HERO is set up to handle frequency dependent opacities.  Both the computational cost and memory requirement scale linearly with $n_F$ the number of frequency bins.  All radiative quantities (opacities, intensities, source functions) are calculated for the same discrete set of frequencies.  The code could be set up to handle group mean opacities and emissivities with their appropriate quadrature weights to handle more complex line emission problems.  In problems where there is frequency coupling (i.e. gravitational/doppler shifting) the redistribution of photons is handled by linear interpolation.  Our typical choice for the frequency grid is a total range of 6 decades with a resolution of 10 points per decade (60 frequency bins).

\subsection{Angular Discretization}\label{sec:angleInterp}

The angular setup of our code differs based on the dimensionality of the problem that is being solved.

\subsubsection{1D Angles}
In 1D, we solve plane parallel slab problems which are actually 3D problems that can be collapsed down to 1D by invoking translational and azimuthal symmetry.  We subdivide the $4\pi$ steradian sphere into equal solid angle slices in $\theta$; thus, we set our angle spacing $d\theta$ such that $\sin\theta d\theta = 2/N_A$.

\subsubsection{2D Angles}
In all the 2D test problems described in this paper, the rays are considered to live in a flat 2D space, so we do not consider the 3D solid angle at all.  We choose our angle grid so as to cover uniformly the full $2\pi$ of the 2D plane.  We use simple linear interpolation on 2D angles to handle the mixing of angles when rays travel from one cell to another.

\subsubsection{3D Angles}\label{sec:3Dangles}

The goal is to subdivide the $4\pi$ sphere of solid angle into $N_A$ solid angle wedges.  A few common approaches include bisecting octants, following a "spiral" that winds around the sphere from the two poles, or a special grid of latitudes/longitudes. All these methods approximately achieve the required goal, but they suffer from undesirable patterns of symmetry lines or "seams" on the surface of the sphere.  To sidestep this issue, we obtain our angular grid via a numerical approach.

\begin{figure}
\begin{center}  
\subfigure{\includegraphics[width=0.3\textwidth]{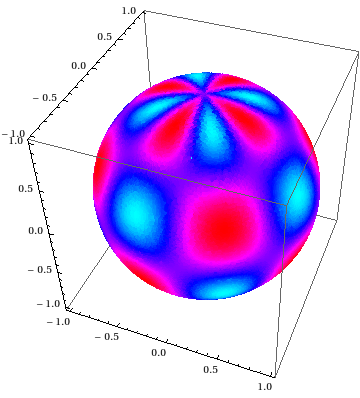}}
\subfigure{\includegraphics[width=0.3\textwidth]{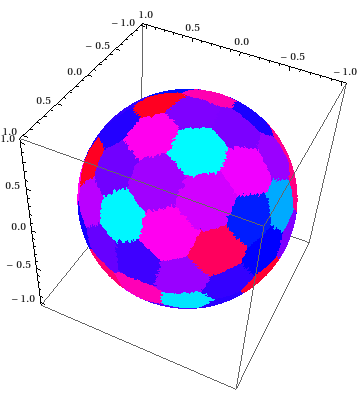}}
\subfigure{\includegraphics[width=0.3\textwidth]{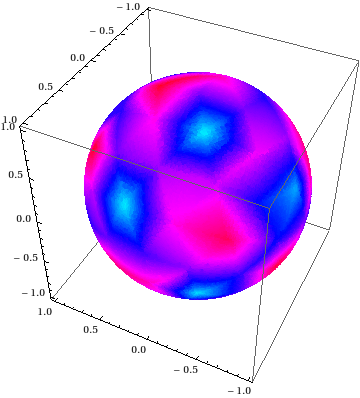}}
\caption{A plot of how our interpolation scheme performs compared to the exact solution.  Top: Exact $I = \cos(4\theta)\cos(4\phi)$ beam pattern, Middle: discretized version of top panel using our $N_A=80$ predefined grid, Bottom: $I$ generated from linear interpolation of the $N_A=80$ grid.
\label{fig:angleInterp}}
\end{center}  
\end{figure}

The strategy that we employ is to deposit $N_A$ fictitious charged particles constrained to move on the surface of a sphere, with initial positions set by simple heuristics (equal proper length spacings in $\theta, \phi$ coordinates).  Then, we allow the particles to evolve over time under their mutual inter-particle electrostatic forces until they settle on an equilibrium configuration.  This naturally produces a set of positions that are nearly equidistant from one another. We use these positions to specify the angular grid (see Fig \ref{fig:angleInterp} for an example with $N_A=80$).

For relativistic problems where light bending introduces mixing between different angle bins, or for curvilinear coordinate grid setups that invoke angle grids fixed along the locally rotated unit direction vectors, we handle angular interpolation via linear combinations.  For a ray with a given unit direction vector $\hat{\psi}$, there will always by a set of 3 angular points ($\psi_1, \psi_2, \psi_3$) on the $4\pi$ sphere which defines a triangle enclosing $\hat{\psi}$.  We find the angles ($\psi_1, \psi_2, \psi_3$) by simply locating the 3 angles with the largest dot product to $\hat{\psi}$ in the angle grid.

Once the three vertices of the triangle have been identified, we decompose $\hat{\psi}$ as a linear combination of $\psi_1$, $\psi_2$, $\psi_3$:
\begin{equation}
\hat{\psi} = c_1 \psi_1 + c_2 \psi_2 + c_3 \psi_3. \label{eq:angleInterp}
\end{equation}
The coefficients $c_1, c_2, c_3$ (after renormalizing to satisfy $c_1 + c_2 + c_3 = 1$) are then used as the interpolation weights.  In rare cases (i.e. when $\psi$ is slightly outside the boundary of the triangle formed by $\psi_1, \psi_2, \psi_3$), the interpolation weights may be slightly negative.  For stability purposes, we clip any negative weights to zero.

In Fig. \ref{fig:angleInterp}, we show how our linear interpolation scheme handles an example intensity pattern.  Notice that in the slowly-varying equatorial regions, the interpolated result provides a much better match to the exact intensity distribution than simply using a discretized version of the intensity field.  The linear reconstruction does poorly when the radiation field fluctuates on angular scales smaller than a single beamwidth of the $N_A=80$ grid (e.g. see polar regions of Fig. \ref{fig:angleInterp}).

%


\section{Numerical Tests}\label{sec:results}

We wish to verify the following properties of HERO: 1) that it correctly solves for the radiation field using both short and long characteristics, 2) that it calculates a self consistent gas temperature taking into account the condition of radiative equilibrium, and 3) that it correctly treats the curved space-time of GR.  

In all the following examples, we run HERO decoupled from any hydrodynamics.  HERO is a postprocessor to compute the radiation field $I_\nu$ and, if required, a self-consistent temperature $T_{gas}$, given the optical properties (opacities, emission, absorption) of the medium.  Other properties of the background fluid, specifically density and velocity, are assumed to be given and fixed. It is also assumed that the system is time-independent, so we effectively solve for the steady state solution for the radiation.

We are particularly interested in problems where scattering dominates the opacity (as is often true with relativistic accretion flows), leading to diluted-blackbody radiation fields.  Due to the complicated nature of scattering, many iterations are needed for the radiative solver to converge.  This makes scattering problems ideally suited to test the convergence properties of the code. We describe below a series of 1D, 2D and 3D tests.



\subsection{1D Plane-parallel Grey Atmosphere}\label{sec:1DTests}

Scattering problems are in general very difficult to solve due to the integro-differential nature of the scattering kernel in the radiative transfer equation (see righthand side of Eq. \ref{eq:RTstart}).  As such, there are very few examples that have closed form analytic solutions.

The 1D plane-parallel isothermal slab with grey opacity turns out to be one particularly simple case with a well known closed form solution in the two-stream limit (c.f. \S1 \citealt{rybicki79}).  There is also a semi analytic solution in the limit of infinite angles. For this problem, all quantities, $T,\,\kappa,\,\sigma$, are uniform within the medium, leading to a solution that is only a function of the optical depth: $d\tau = (\kappa + \sigma) dz$. Under the two-stream approximation, and using the Eddington approximation to close the radiative moment equations, the analytic solution for the mean intensity is given by
\begin{equation}
J(\tau) = B\left[1 - \frac{\exp(-\sqrt{3\epsilon \tau})}{1+\sqrt{\epsilon}}\right], \label{eq:analytic2stream}
\end{equation}
where $B$ is the thermal blackbody source function, $\tau$ is the total optical depth from the surface, and $\epsilon$ is the photon interaction destruction probability given by Eq.\ref{eq:epsilon}.

\begin{figure}
\begin{center}  
\includegraphics[width=0.45\textwidth]{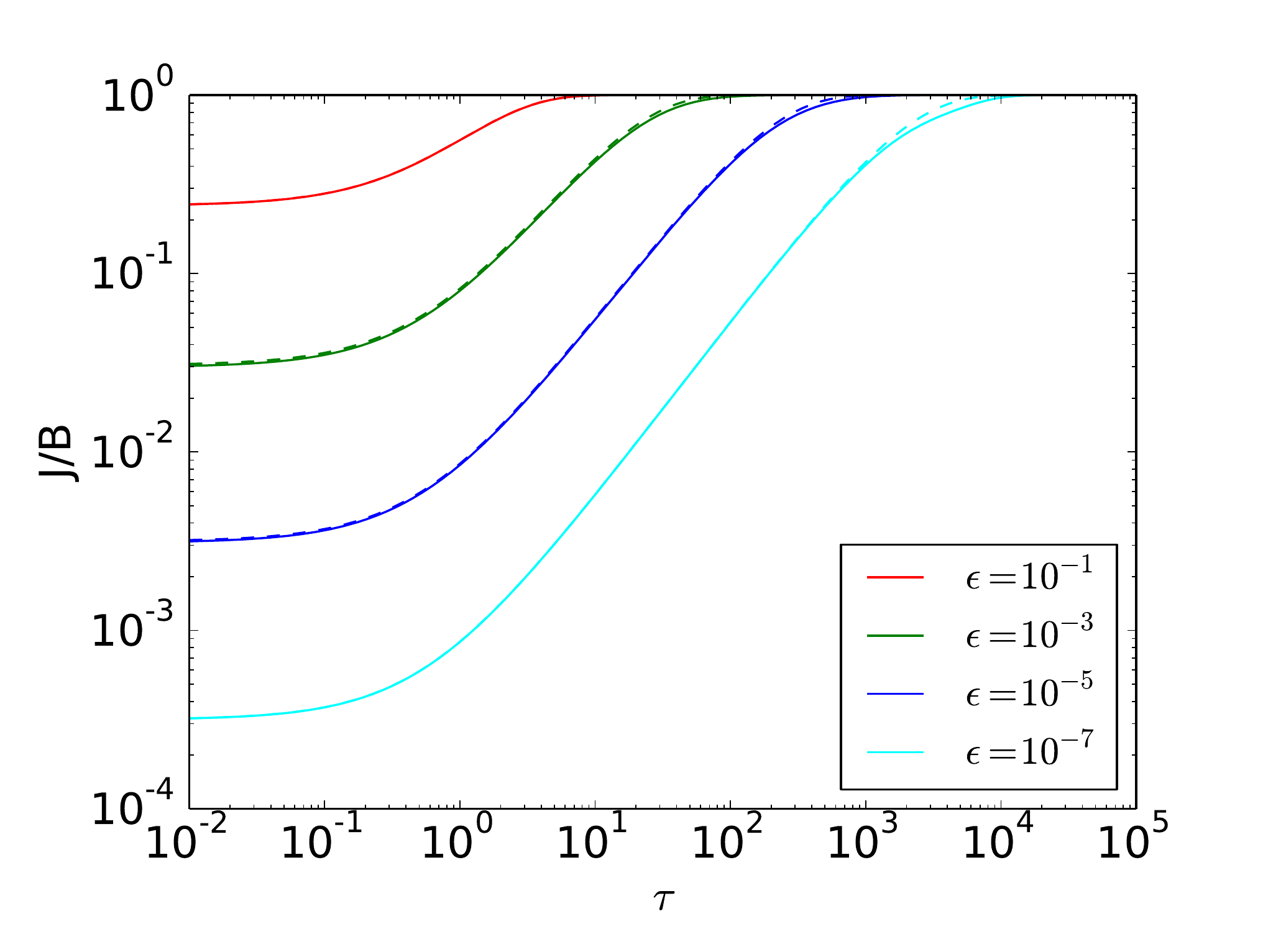}
\caption{$J$ computed from HERO (dashed lines) compared with the analytic solution (solid lines) for a plane parallel  scattering atmosphere described under the 2-stream approximation.  Four values of $\epsilon$ (see Eq. \ref{eq:epsilon}) are shown.
\label{fig:1D-analytic}}
\end{center}  
\end{figure}

Note the exponential transition in $\tau$ that separates a thermal interior from a dilute-blackbody surface layer.  In Figure \ref{fig:1D-analytic}, we show solutions computed by HERO compared to Eq. \ref{eq:analytic2stream}.  The $\tau=0$ surface outer boundary condition was set to have zero ingoing flux, and the innermost $\tau=10^5$ boundary was set to $J=B$ for some fixed $B$.  The agreement between the numerical and analytical solutions is quite good ($\sim1\%$ errors) with a systematic trend of larger errors in low $\epsilon$ systems.  These low $\epsilon$ atmospheres are highly scattering dominated and the induced systematic error is simply a consequence of the nonlocal nature of scattered radiation.

\subsection{Convergence Tests}\label{sec:convergence}
\begin{figure}
\begin{center}  
\includegraphics[width=0.45\textwidth]{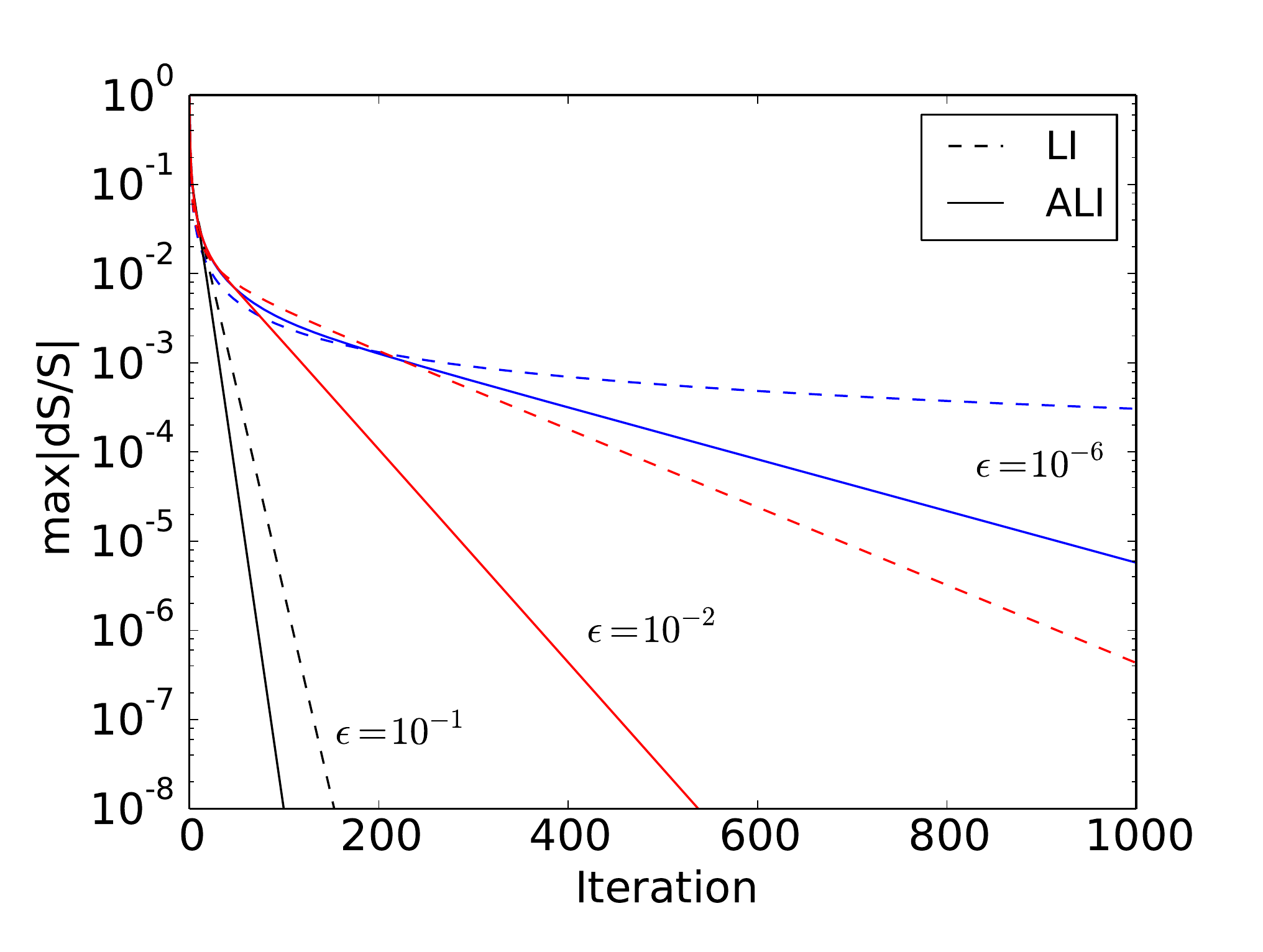}
\caption{Convergence rates for highly scattering plane parallel atmospheres.  Note that the standard Lambda Iteration (LI) procedure saturates at a fixed error threshold for small values of $\epsilon$ (when $\epsilon \ll 10^{-2}$).  Accelerated Lambda Iteration (ALI) using Eq. \ref{eq:ALI} converges for all values of $\epsilon$, even extremely small values.
\label{fig:convergence}}
\end{center}  
\end{figure}

The previous plane-parallel atmosphere calculations for the high scattering cases ($\epsilon < 10^{-3}$)  required ALI to converge.  In Figure \ref{fig:convergence}, We show the convergence properties of the code for a few example problems to emphasize the importance of acceleration in obtaining the correct solution within a reasonable number of iterations.

In the absence of acceleration, radiation information can only propagate a distance $\delta \tau \sim 1$ per iteration.  Thus, the number of iterations needed for the radiation to pass from one end to the other is $\sim \tau_{tot}^2$ iterations (corresponding to random walk diffusion with stepsize $\delta \tau = 1$).

The power of ALI rests in the fact that it boosts the single iteration range of influence from $\delta \tau = 1$ to $\delta \tau = \delta \tau_{cell}$.  To see why this occurs, consider a tiny perturbation of the source function $\Delta S$ for some cell.  The effect of $\Delta S$ on a neighbouring cell is attenuated by an exponential optical depth factor $\exp(-\Delta \tau)$, where $\Delta \tau$ represents the inter-cell optical depth.  The ALI boost factor (c.f. Eq. \ref{eq:ALI} when $\epsilon \rightarrow 1$) in the scattering dominated limit is $\Delta S_{\rm{ALI}} = \Delta S \exp(\Delta \tau)$, which exactly compensates for the $\exp(-\Delta \tau)$ attenuation. This allows the $\Delta S$ information to propagate to the next cell in just a single iteration, hence requiring $N_{cell}$ iterations for information to traverse the medium from end to end.  For most scattering problems, this leads to a huge speedup (since $N_{cells} \ll \tau_{tot}^2$).

\subsection{Multiray Temperature Solution}

\begin{figure}
\begin{center}  
\subfigure{\includegraphics[width=0.45\textwidth]{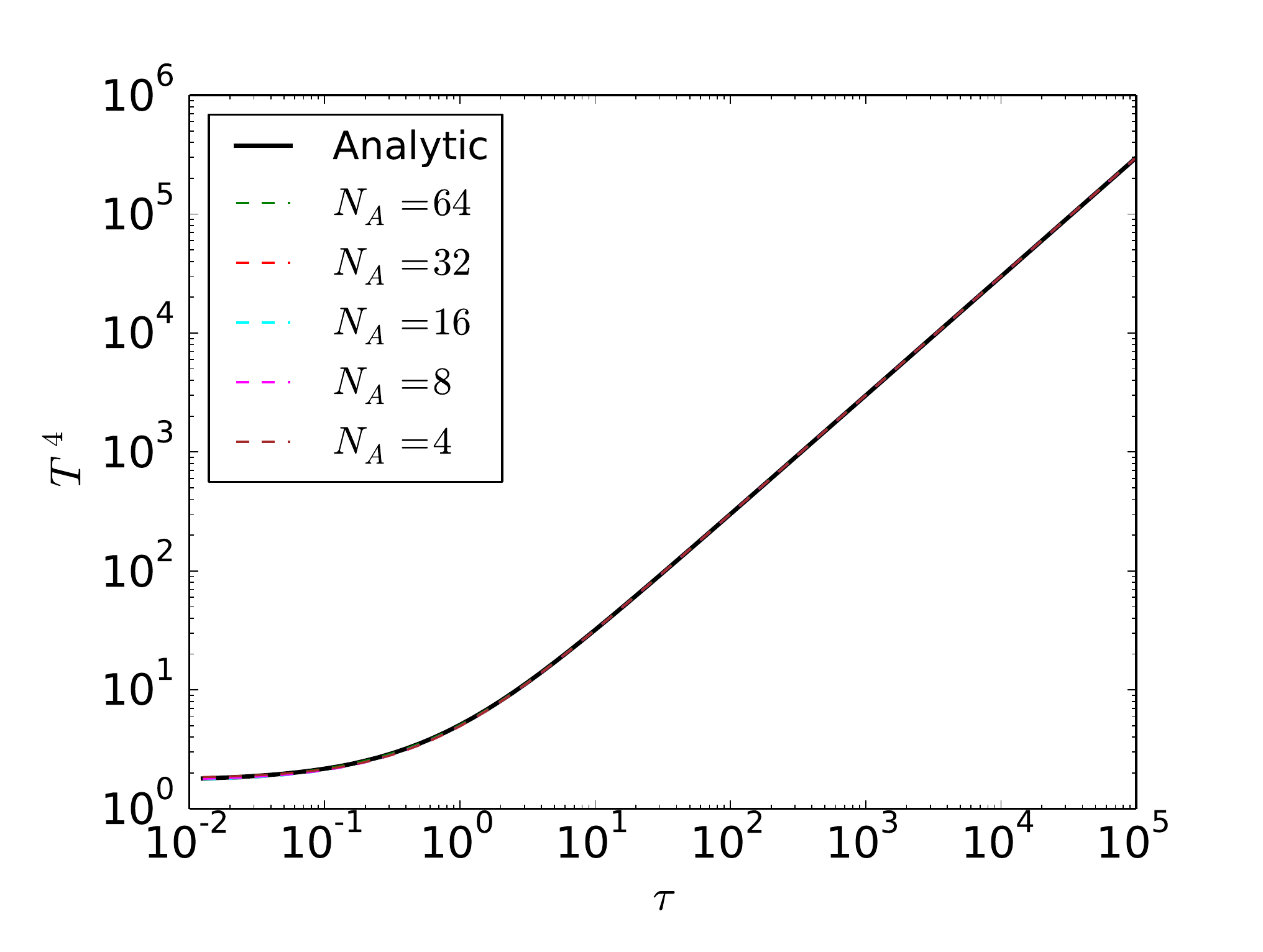}}
\subfigure{\includegraphics[width=0.45\textwidth]{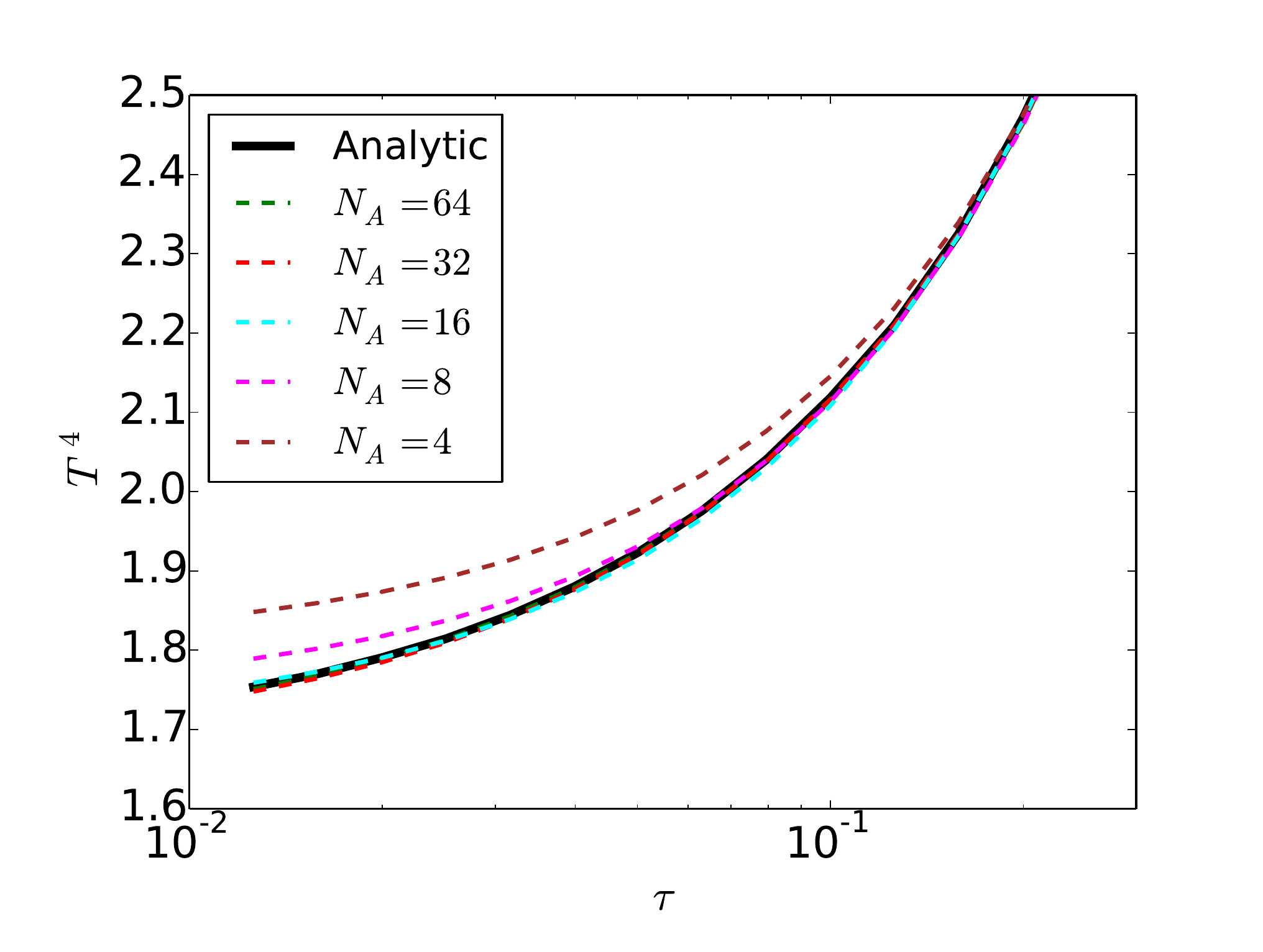}}

\caption{Comparison of analytic solution (solid line) to numerical results (dashed lines) for $\epsilon=1$ constant flux atmospheres modeled with different numbers of angles.  The bottom panel is a zoomed in version of the top panel.  The temperature solution in the interior ($\tau \gg 1$) matches perfectly to the analytic solution.  The agreement at the surface improves as the number of rays is increased.\label{fig:Tsol}}
\end{center}  
\end{figure}

Here we describe a test of HERO's ability to calculate equilibrium temperatures. In Figure \ref{fig:Tsol}, we compute a constant flux nonscattering atmosphere and compare the numerical temperature profile with the analytic result (from a table of the Hopf function, see \citealt{chandrasekhar50}).  With sufficient angular resolution ($N_A > 16$), the temperature profile computed with HERO is correct to within $1\%$.\\[0.5in]

\subsection{Test of Spectral Hardening}

As a test of HERO's multifrequency capabilities, we now examine a frequency dependent 1D atmosphere.  This problem is simple enough to admit an exact analytic solution for the spectrum at all depths within the atmosphere.  Appendix \ref{app:analyticSpectrum} derives the depth dependent spectrum.

We are particularly interested in examining how a strongly scattering atmosphere modifies the thermal emission from deep within a plane parallel atmosphere.  This phenomenon is known as spectral hardening, and acts to shift the apparent colour of the photosphere emission.  The effect is particularly important in the context of modelling black hole accretion disc spectra \citep{shimura95,davis06}, where the high degree of scattering leads to a shift in the thermal emission peak by a factor of $\sim1.5-2$ towards higher energies.

\begin{figure}
\begin{center}
\includegraphics[width=0.5\textwidth]{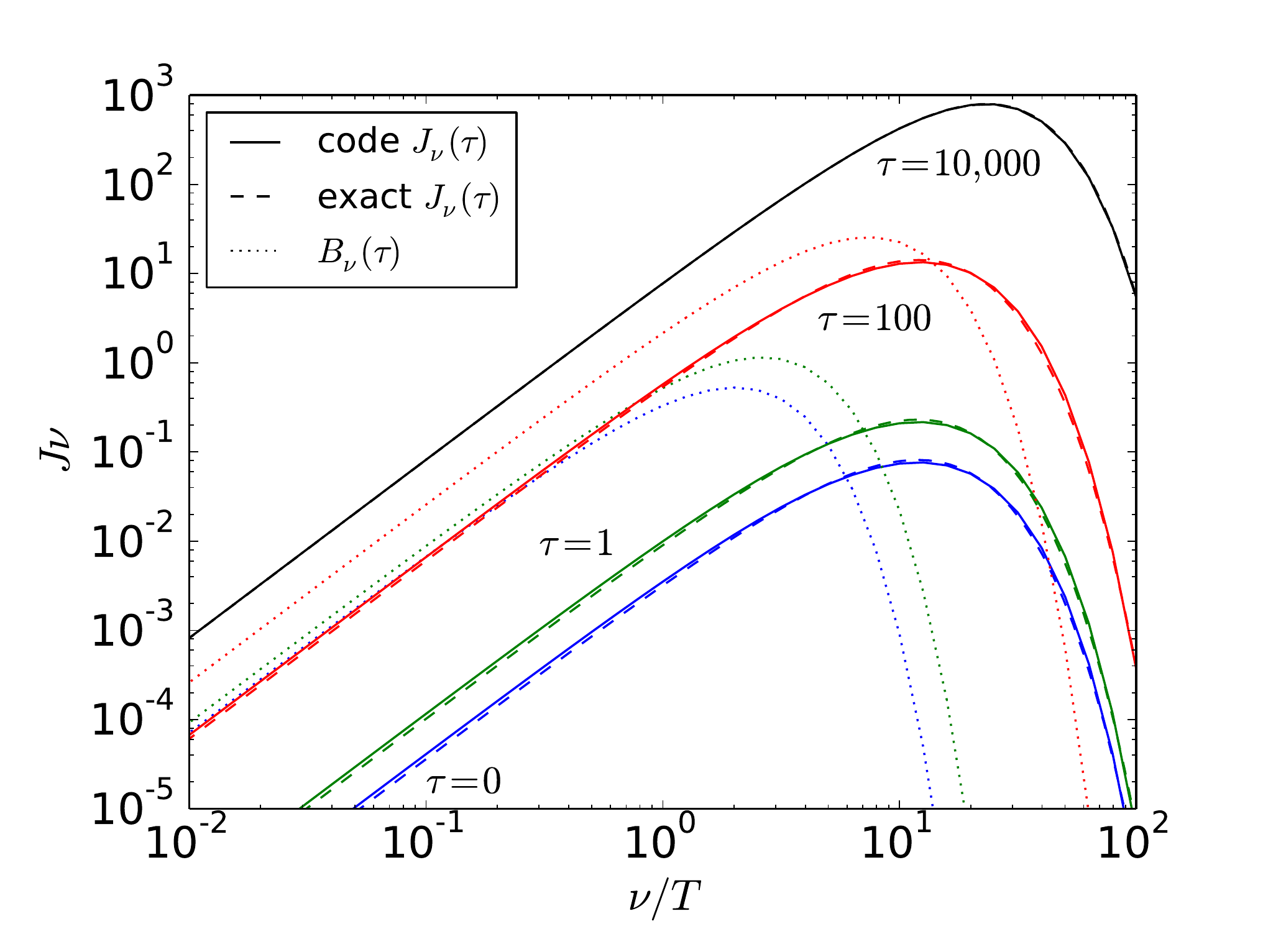}
\caption{Comparison of depth dependent spectra calculated using HERO (solid lines) and the analytic solution (dashed lines) for a plane parallel thermal atmosphere.  Note: this is a \emph{highly} scattering test problem with $\epsilon=10^{-6}$, which results in significant spectral hardening at the $\tau=0$ surface (compare dotted blue blackbody curve with solid blue line).  At all depths and frequencies, HERO is able to calculate the spectrum to within $5\%$ of the true solution.  The spatial resolution chosen for this calculation was 25 points per decade in $\tau$, and 20 points per decade in $\nu$. The two-stream approximation was used for handling angles ($N_A=2$).
\label{fig:spectest}}
\end{center}
\end{figure}

Figure \ref{fig:spectest} demonstrates an extreme version of this spectral hardening effect for an $\epsilon = 10^{-6}$ plane-parallel medium.  HERO produces spectra that agree to within 5\% of the analytic solution for all energies and optical depths.  Note that above the photosphere ($\tau < \tau_{\rm eff} = 1/\sqrt{3\epsilon} \approx 10^3$), the radiation spectrum differs quite substantially from the local thermal emissivity/temperature (dotted curves).

For simplicity, the example shown here is with the two-stream approximation for grey (i.e. frequency independent) homogeneous opacities.  We impose a constant flux boundary condition at $\tau = 10^5$ and solve numerically for the equilibrium temperature and frequency-dependent radiation intensity as a function of depth.  There is excellent agreement between the code output and the analytic temperature and radiation profiles (Eq. \ref{eq:spectralTSoln} in Appendix B).

\subsection{Effect of a Heating Source}

As a final accretion disc-like 1D test that utilizes most of the features of the code, we discuss a problem that includes a heating source. We consider a slab with a total optical depth of $2\times10^5$ between two free surfaces. We assume grey opacities with $\epsilon = 10^{-4}$ (a strongly scattering-dominated disc). The material in the slab is heated at a steady rate of $10^{-5}$ (arbitrary units) per unit optical depth, so that the steady state flux from each surface is unity. We use 20 angles and 61 frequencies.

\begin{figure}
\begin{center}  
\subfigure{\includegraphics[width=0.45\textwidth]{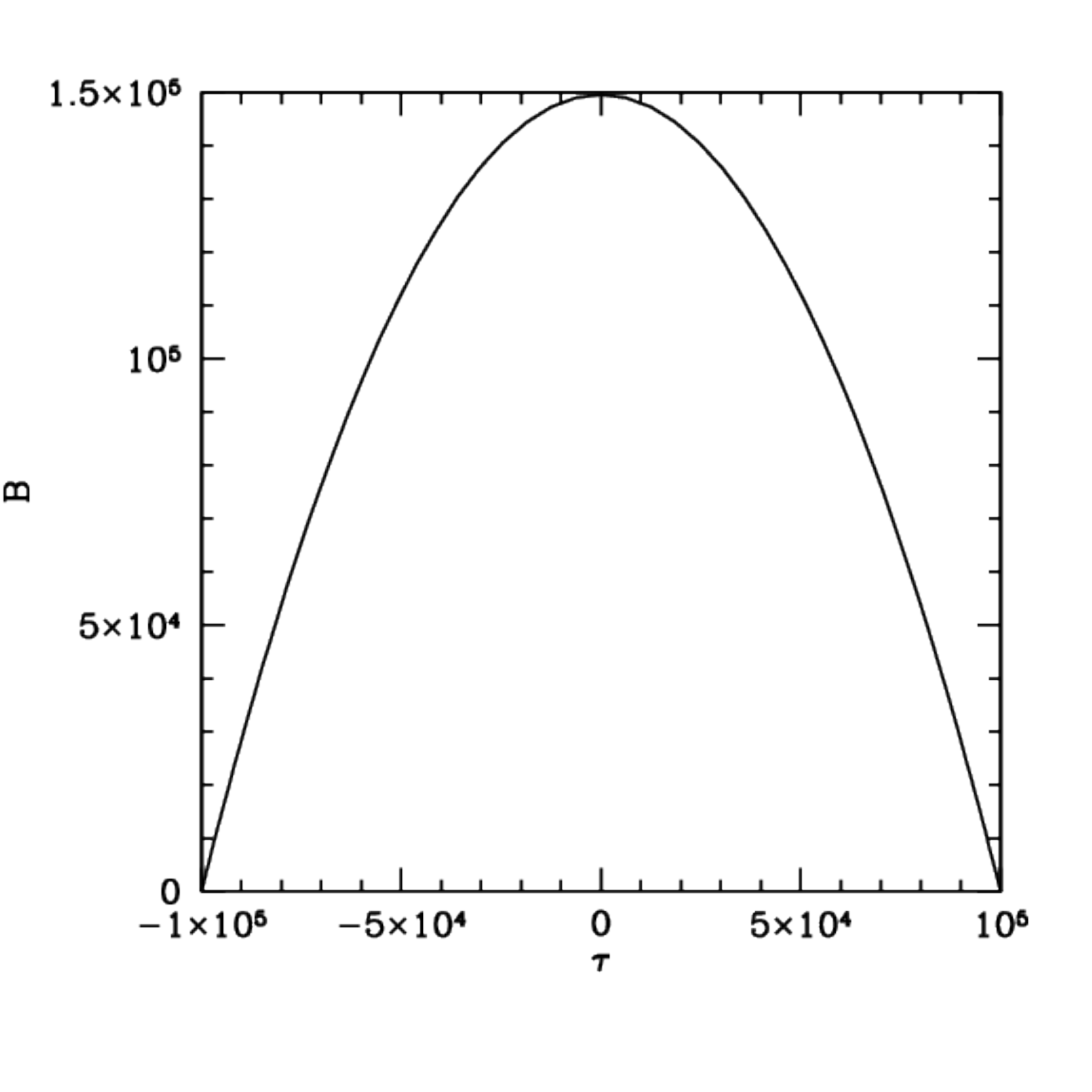}}
\vskip -1.5cm
\subfigure{\includegraphics[width=0.45\textwidth]{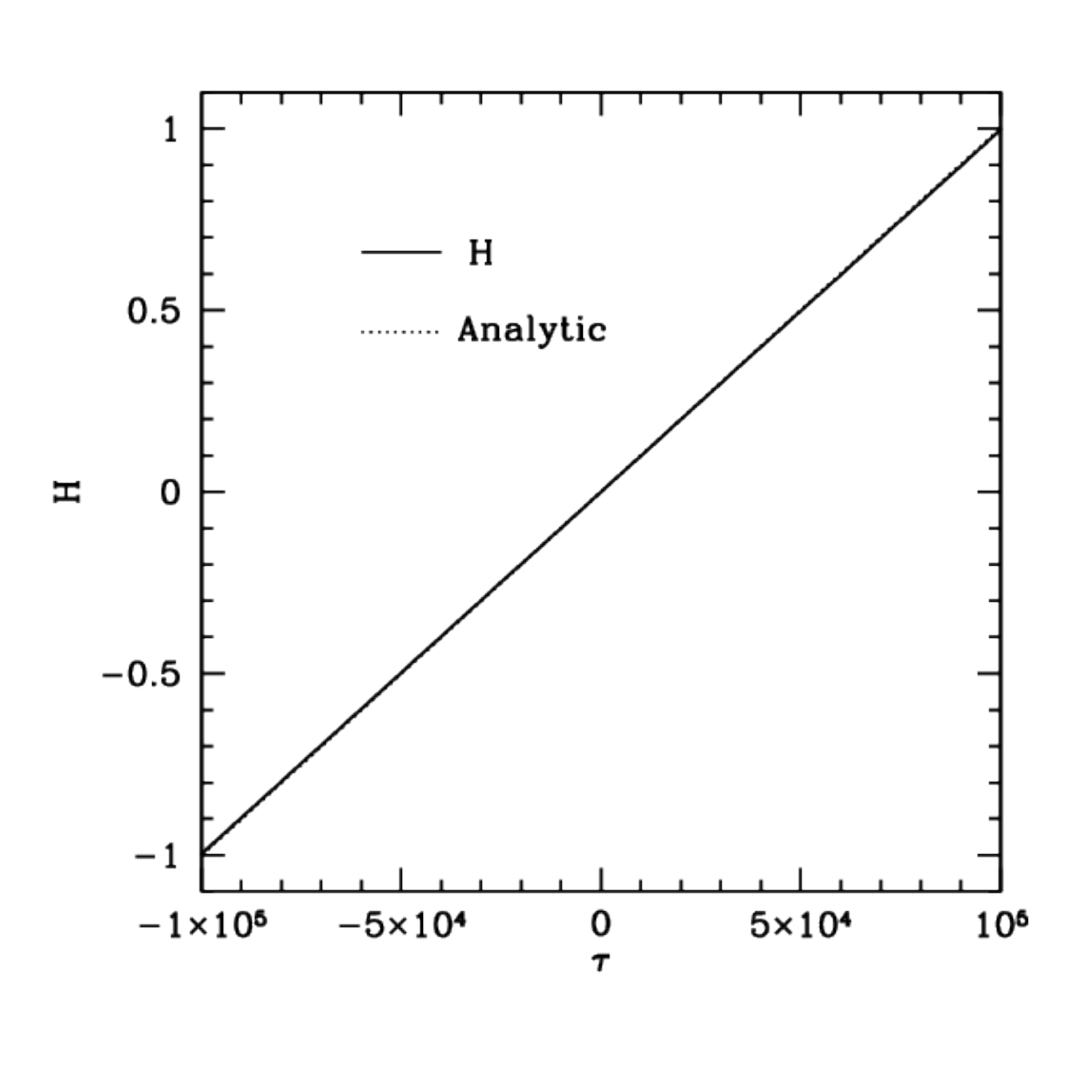}}
\vskip -1.5cm
\subfigure{\includegraphics[width=0.45\textwidth]{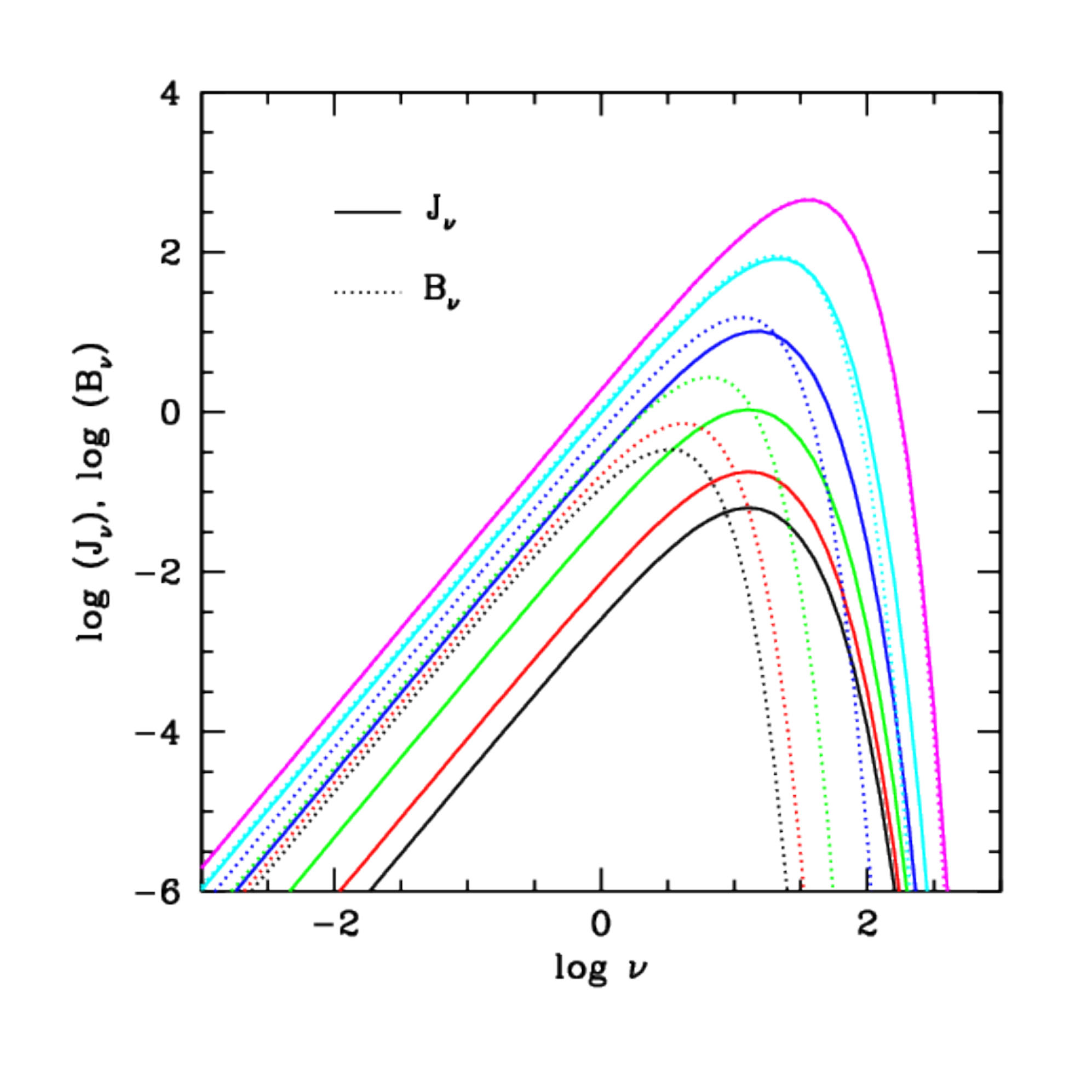}}
\vskip -1.cm
\caption{Top: Self-consistent solution with HERO for the temperature in the interior of a uniformly heated 1D slab.  Middle: Numerical solution for the frequency integrated radiative flux $H$ compared with the analytic solution.  Bottom: Radiation spectrum $J_\nu$ compared with the blackbody spectrum at the local temperature $B_\nu$ for optical depths (from below) $0,\,1,\,10,\,10^2,\,10^3,\,10^4$ from the surface.
\label{fig:heating}}
\end{center}  
\end{figure}

We start the calculation with some arbitrary temperature profile ($B = 10^4$ at all $\tau$ in this particular example) and we run HERO until it converges. After 3000 iterations (this is many more than needed, but we wished to converge as much as possible for this test), the radiative transfer equation has a maximum error of $10^{-6}$ and the radiative equilibrium equation a maximum error of $10^{-7}$.  The results are shown in Fig.~\ref{fig:heating}. As expected, the temperature profile is quadratic in $\tau$ and symmetric with respect to the mid-plane (upper panel) and the radiation flux is linear in $\tau$ and antisymmetric (middle panel). The lower panel shows spectra at various depths. In the deep interior of the disc, the radiation is nearly blackbody, but as we approach the surface ($\tau_{\rm eff} = \tau \sqrt{3\epsilon} < 1$) there are signatures of spectral hardening. The radiation that escapes from the surface is distinctly hardened.

\subsection{2D Solutions and Ray Defects}

A well known limitation of radiative solvers with discretized angular zones is the appearance of ``ray defects''.  These are sharp linear features that arise because interpolation on the angular grid is imperfect.  The presence of ray defects is an important motivator for us to use a hybrid scheme.  Our long characteristics method operates independently of an angular grid since it adaptively chooses an angular grid and hence does not suffer from ray-defects.  In the following sections, we compare our two solvers, SC and LC, for a few test 2D problems.  Appendix \ref{app:raydefects} provides a more extensive discussion of the ray defects.

\subsubsection{Opaque Wall Test}

\begin{figure}
\begin{center}  
\subfigure{\includegraphics[width=0.5\textwidth]{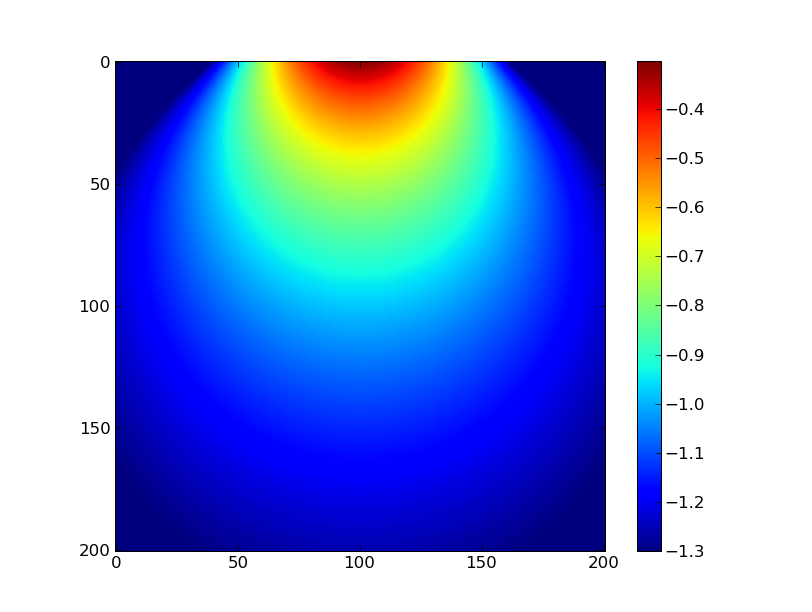}}
\subfigure{\includegraphics[width=0.5\textwidth]{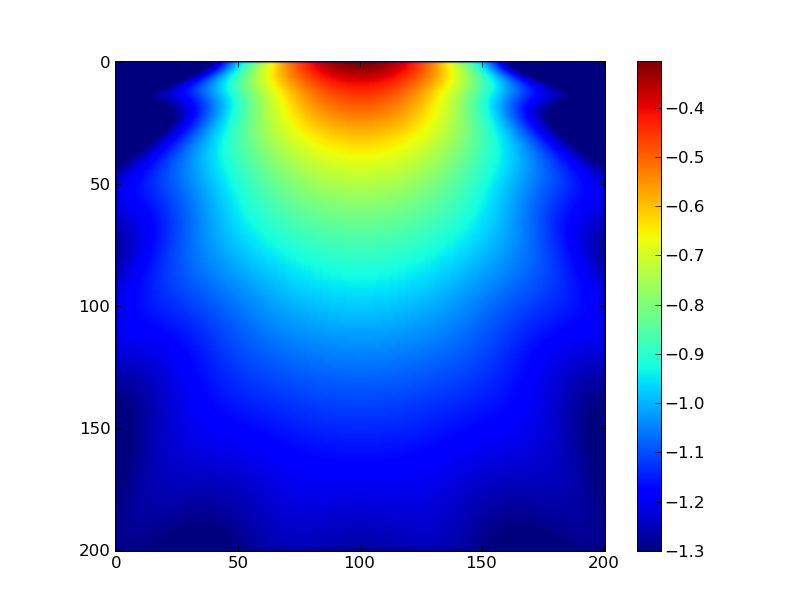}}
\subfigure{\includegraphics[width=0.5\textwidth]{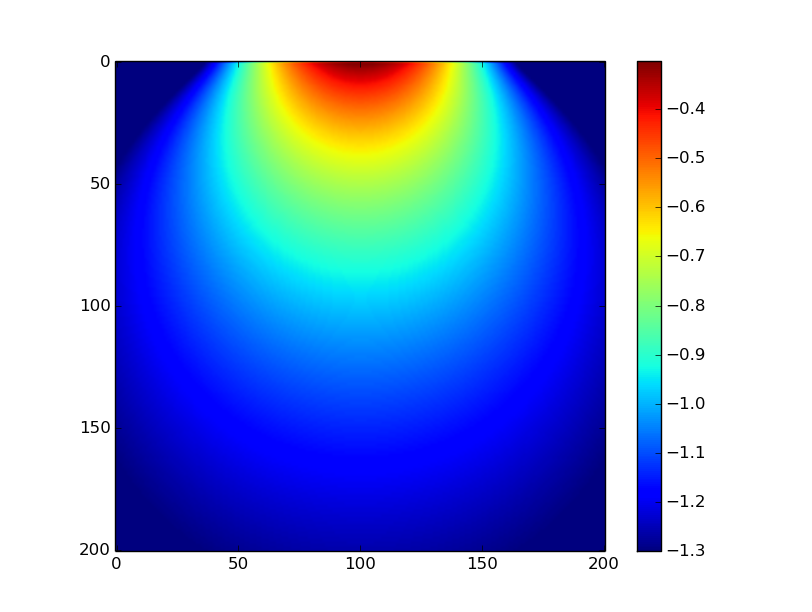}}
\caption{Comparison of radiation fields calculated using short and long characteristics with an exact analytic solution for an opaque gaussian wall emitting into vacuum (colour $\log J$).  From top to bottom:  a) exact solution, b) SC solution ($N_A = 20$), c) LC solution.  Note the appearance of systematic banding/waves in the SC case. These are ray defects.
\label{fig:2D-analytic}}
\end{center}  
\end{figure}

Ray-defects are particularly severe when there is a compact source of radiation (see Fig. \ref{fig:ray-defects} in Appendix \ref{app:raydefects}). But there are noticeable effects even in the case of smooth extended sources.  As an illustration, we analyze a simple 2D example where it is easy to compute the exact solution.  In the top panel of Figure \ref{fig:2D-analytic}, we show the radiation field for an empty (zero opacity) box illuminated by a hot spot in the centre of the upper wall.  The boundary condition at this wall is an opaque ($\tau=\infty$) isotropic source function with a gaussian profile:
\begin{equation}
\frac{S(x)}{S(x_c)} = \exp\left[-\frac{(x-x_c)^2}{w}\right],
\end{equation}
where $x_c$ corresponds to the centre of the box, and the width $w$ is equal to $1/5$ of the box size.  Since the interior of the box has no opacity, the radiation field at any point can be easily found by tracing rays backwards and finding the  source function corresponding to the ray's intersection point with the upper wall.  We can thus calculate the mean intensity $J$ at every point inside the box.

In Figure \ref{fig:2D-analytic}, we compare the exact solution to the result from our two RT solvers.  For the SC solver (middle panel), a clear wave pattern appears along each of the angle grid directions (we used $N_A=20$ in this test). These are the ray defects mentioned earlier.  The severity of the defects grows as we approach the upper wall, because of the sharp break in the intensity distribution there.  On the other hand, LC perfectly reproduces the radiation pattern, even in the far field limit (compare top and bottom panels of \ref{fig:2D-analytic}).

\subsubsection{Shadowing Test}
Shadowing is another classic test that is useful for spotting systematic biases in a radiative solver.  Simple moment closure schemes such as FLD \citep{levermore81} or M1 \citep{levermore84} typically have problems resolving the correct shadow structure.  For instance, FLD has trouble maintaining the coherency of shadows over long distances due to its diffusive nature, and M1 has issues dealing with multiple light sources.

\begin{figure}
\begin{center}  
\subfigure{\includegraphics[width=0.5\textwidth]{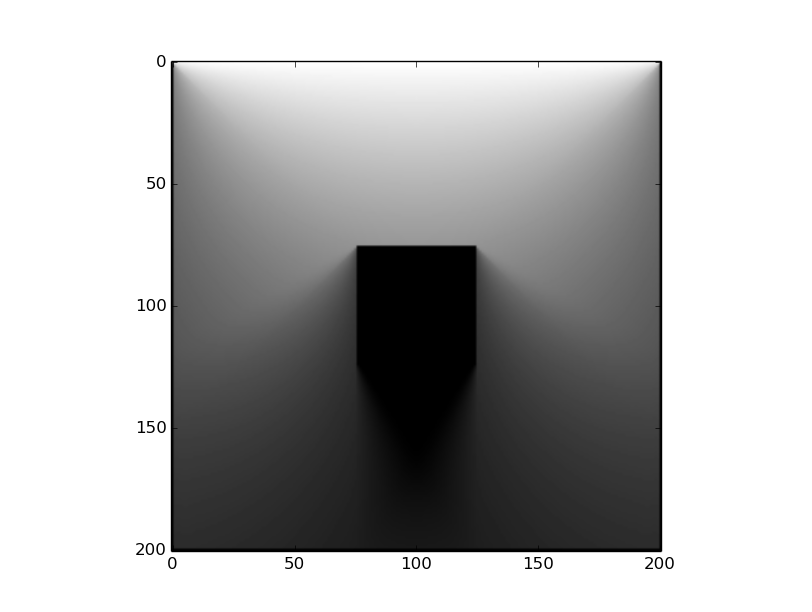}}
\subfigure{\includegraphics[width=0.5\textwidth]{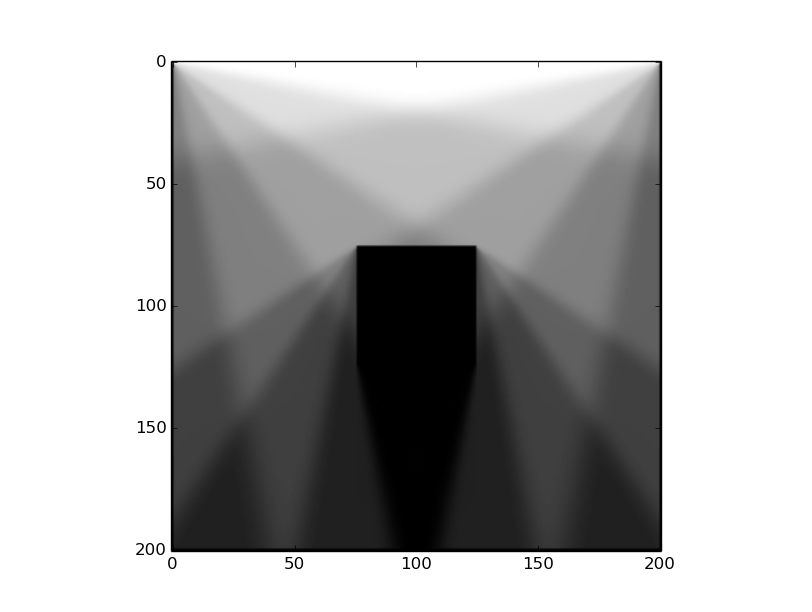}}
\subfigure{\includegraphics[width=0.5\textwidth]{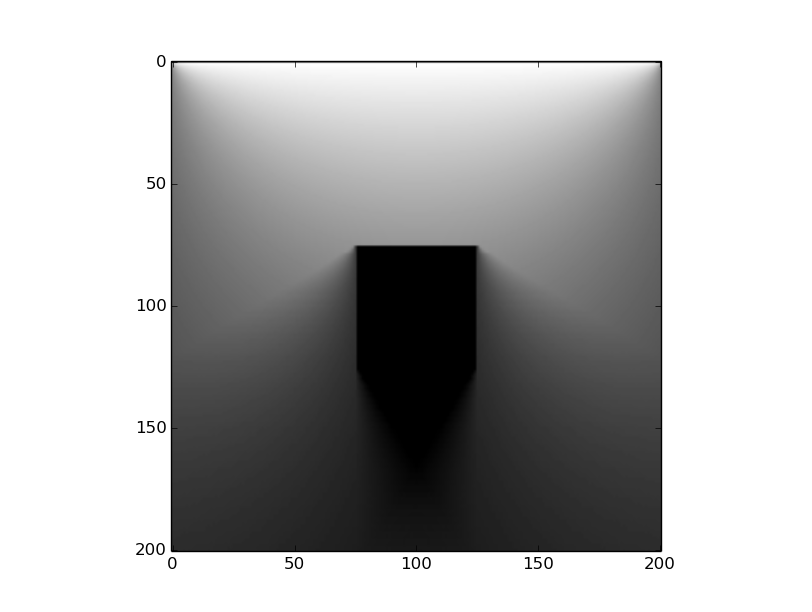}}
\caption{Comparison of shadow patterns calculated using short and long characteristics for an isotropically emitting top wall shining on a central opaque box.  From top to bottom:  a) ``exact" solution ($N_A = 1000$), b) SC solution ($N_A = 16$), c) LC solution.  Note the discrete levels (ray defects) in the SC shadow pattern.
\label{fig:shadow}}
\end{center}  
\end{figure}

In Figure \ref{fig:shadow}, we consider the shadow structure produced by an optically thick square box illuminated from above by an isotropically emitting wall.  The top panel shows the true solution, with a clear umbra and penumbra.  The middle panel shows the result using HERO SC with a crude angular grid $N_A=16$.  Note that the beam resolution of $20^o$ is our typical angular resolution in 3D problems (80 rays in 3D), so this is a realistic example of how SC would perform in 3D.  Increasing the angular resolution makes the ray defects less pronounced -- the top ``exact" solution was produced with $N_A=1000$ using our SC solver.  

The bottom panel shows the result with the LC solver. Notice the high accuracy of LC in handling the radiation shadow pattern.  This is a common theme in all of our tests.  LC is always much superior to SC.
%
\subsection{3D Solutions}

We now shift our attention to 3D test problems, and begin by discussing ray-defects.  All the ray defects discussed in 2D (previous subsection and also Appendix \ref{app:raydefects}) are present in 3D as well, especially in the case of cartesian grids.  A spherical polar grid (the most natural choice for accretion problems) eliminates some problems by introducing ray mixing via the curvilinear nature of the coordinate system. However, this is at the expense of introducing a particularly serious defect for radial rays moving out from a central source. Specifically, if one applies the SC method blindly on a spherical grid, one will obtain a constant radiation energy density and flux at large radius instead of the inverse square law fall-off one expects. 


\citet{dullemond00} discuss a way to ``fix'' the radial beam problem. They slightly modify the radiative transfer equation along the radial direction such that the expected inverse-square falloff is recovered.  We build on their suggestion, except that, instead of modifying only one ray (the radial one), we treat all rays equally and apply an artificial diffusion that, when coupled with a logarithmic radial grid, naturally produces an inverse square falloff in the flux.  Details of our diffusion method are explained in Appendix \S\ref{app:diffusionscheme}.  The following tests as well as those in \S\ref{sec:GRsolutions} employ this ray diffusion scheme.

\subsubsection{Ring Benchmark}\label{sec:ringbenchmark}
One particularly simple test problem is an axisymmetric opaque emitting ring in vacuum, for which it is straightforward to calculate the radiation field at any point in space analytically.  For an infinitesimally thin ring emitting at the equatorial plane, the total light reaching any position $\vec{r}$ is given by a 1-dimensional integral.
\begin{equation}
J(\vec{r}) = \int \frac{C}{|\vec{r} - \vec{r}_{\rm{ring}}(\phi')|^2} d\phi' ,
\end{equation}
where $C$ is a constant specifying the emission per unit length of the ring.  In spherical coordinates, we have
\begin{equation}
J(r, \theta, \phi) = \int \frac{C}{r^2 + r'^2 - 2 r r' \sin\theta \sin\theta'\cos(\phi - \phi')} d\phi',
\end{equation}
where  $r' = 0.5,\,\theta' = \pi/2, \,\phi' \in [0,2\pi]$ are the coordinates of the ring in the setup described here.

\begin{figure*}
\begin{center}  
\subfigure{\includegraphics[width=0.3\textwidth]{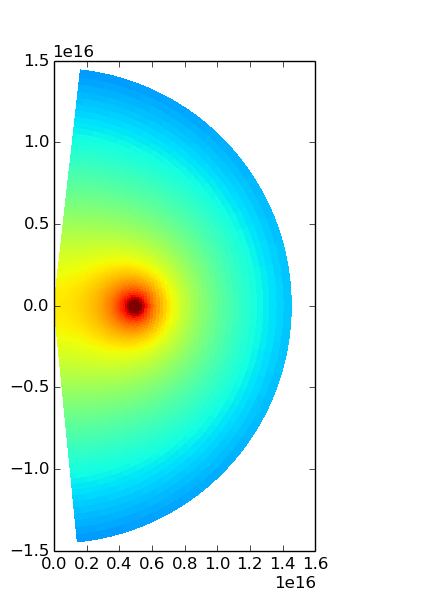}}
\subfigure{\includegraphics[width=0.3\textwidth]{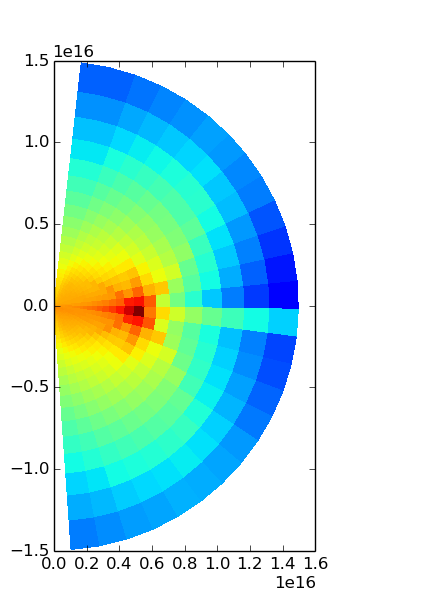}}
\subfigure{\includegraphics[width=0.3\textwidth]{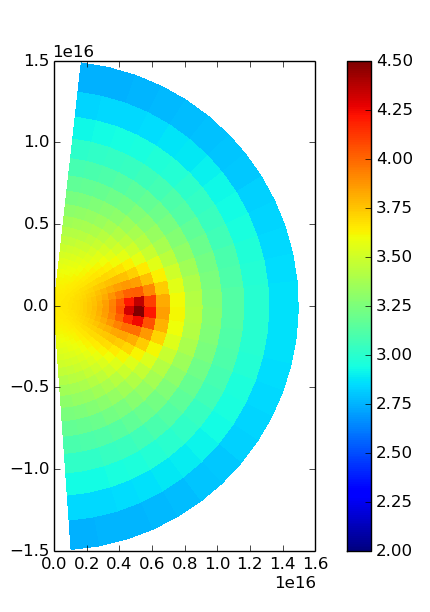}}

\caption{Comparison of true solution (left) with the results obtained with SC (middle) and LC (right) for an axisymmetric emitting ring (colours represent $\log J$).  Note the effect of ray defects  producing a ``spider" pattern for SC.  The LC calculation matches the exact analytic answer to within 0.1\%.
\label{fig:3D-Ring}}
\end{center}  
\end{figure*}

In Figure \ref{fig:3D-Ring}, we compare the results of SC and LC to the analytic result; the constant $C$ has been appropriately normalized to account for the finite emitting area used in the LC/SC calculations.  The LC calculation captures the radiation field perfectly ($\delta J/J \approx 10^{-3}$), whereas the SC result shows strong systematic ray-defect patterns.  Note that to enforce a $1/r^2$ falloff of the radiation field, the SC calculation employs our ray diffusion scheme (otherwise, the result would be significantly worse).

\subsubsection{Dusty Torus Benchmark}
Previously, in \S\ref{sec:1DTests}, we demonstrated that HERO correctly computes both the radiation field and the gas temperature in 1D.  Unfortunately, there are no analytic solutions available for nontrivial 3D problems. Therefore, we turn to a standard benchmark problem that has been widely discussed in the literature and use this problem to numerically compare HERO with other radiative codes.  The model in question consists of an axisymmetric dusty torus \citet{pascucci04} with density structure given by:
\begin{align}
\rho(r,z) &= \rho_0 \cdot f_1(r) \cdot f_2(z)\\ \nonumber
f_1(r) &= (r/r_d)^{-1}\\ \nonumber
f_2(z) &= \exp\{-\pi/4 [z/h(r)]^2\}\\ \nonumber
h(r) &= z_d (r/r_d)^{9/8}
\end{align}
The opacities are tabulated and correspond to $0.12\mu m$ silicate grains \citep{draine84}.  Scattering is assumed to be isotropic, dominating in the wavelength range $0.2 - 1.0\mu m$.  A stellar point source is located at the centre of the disc.  This point source shines on the disc and dictates its energetics and temperature structure.  In HERO, the radiation emanating from the central point source is set to the exact stellar solution at the innermost radial cells of the grid.  The propagation of this radiation outwards is then handled by the short (or long) characteristics solver.  To avoid being killed by severe ray-defects, we include the diffusive term in the short characteristics solver as described earlier (see \S\ref{app:diffusionscheme}). We apply free outflowing radiation boundary conditions at the outer radius of the grid. 

\begin{figure*}
\begin{center}  
\subfigure{\includegraphics[width=0.45\textwidth]{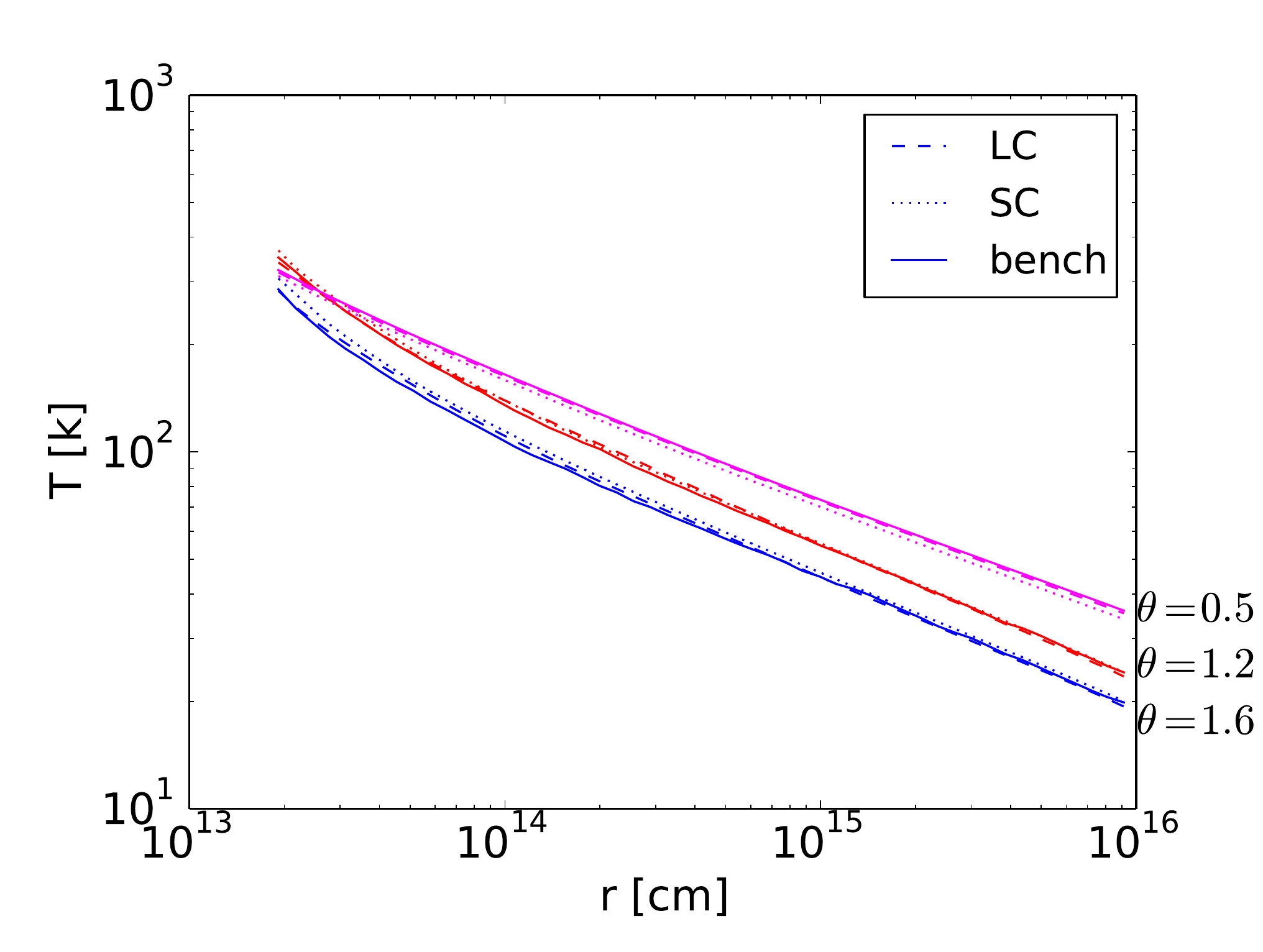}}
\subfigure{\includegraphics[width=0.45\textwidth]{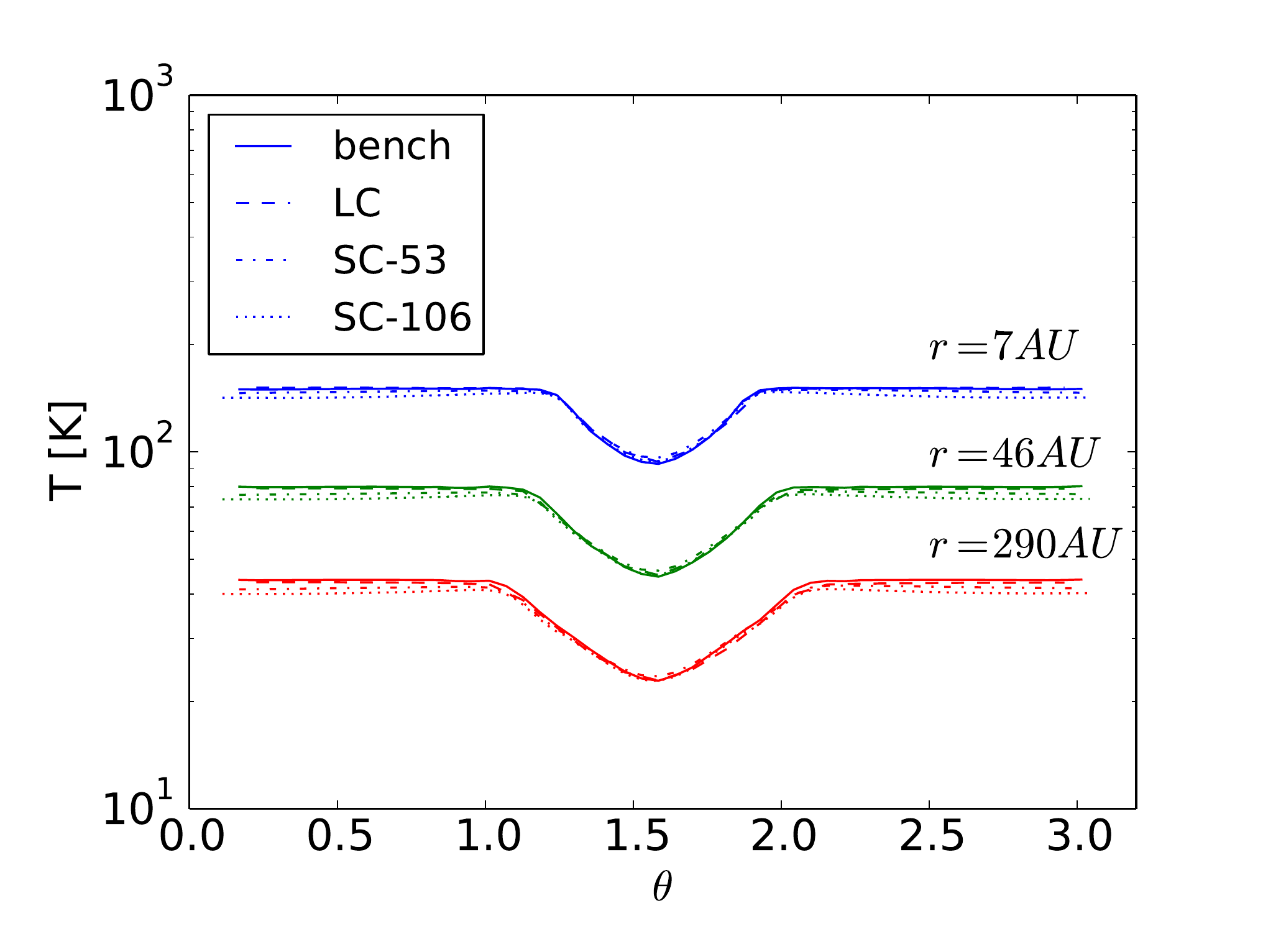}}
\subfigure{\includegraphics[width=0.45\textwidth]{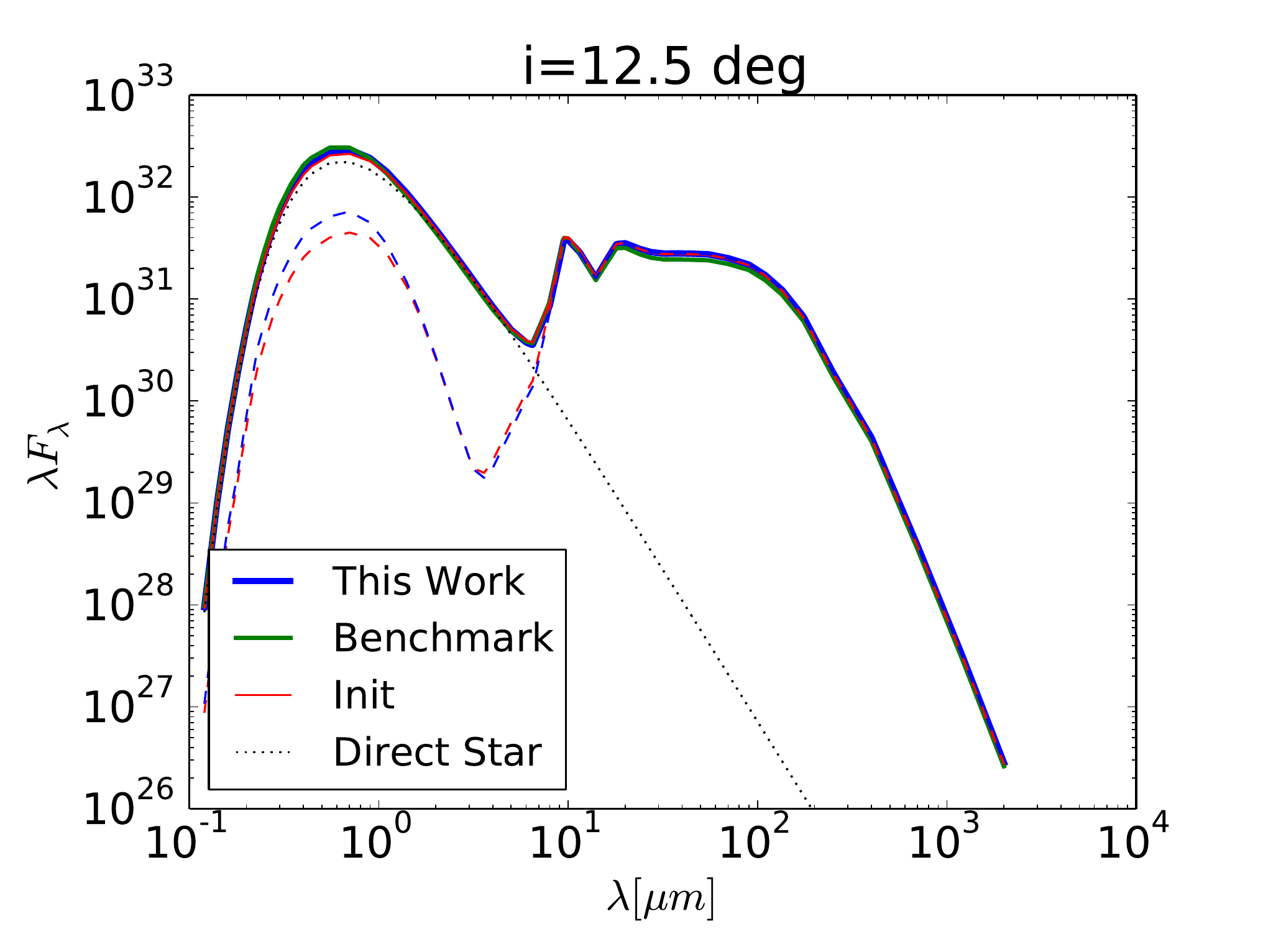}}
\subfigure{\includegraphics[width=0.45\textwidth]{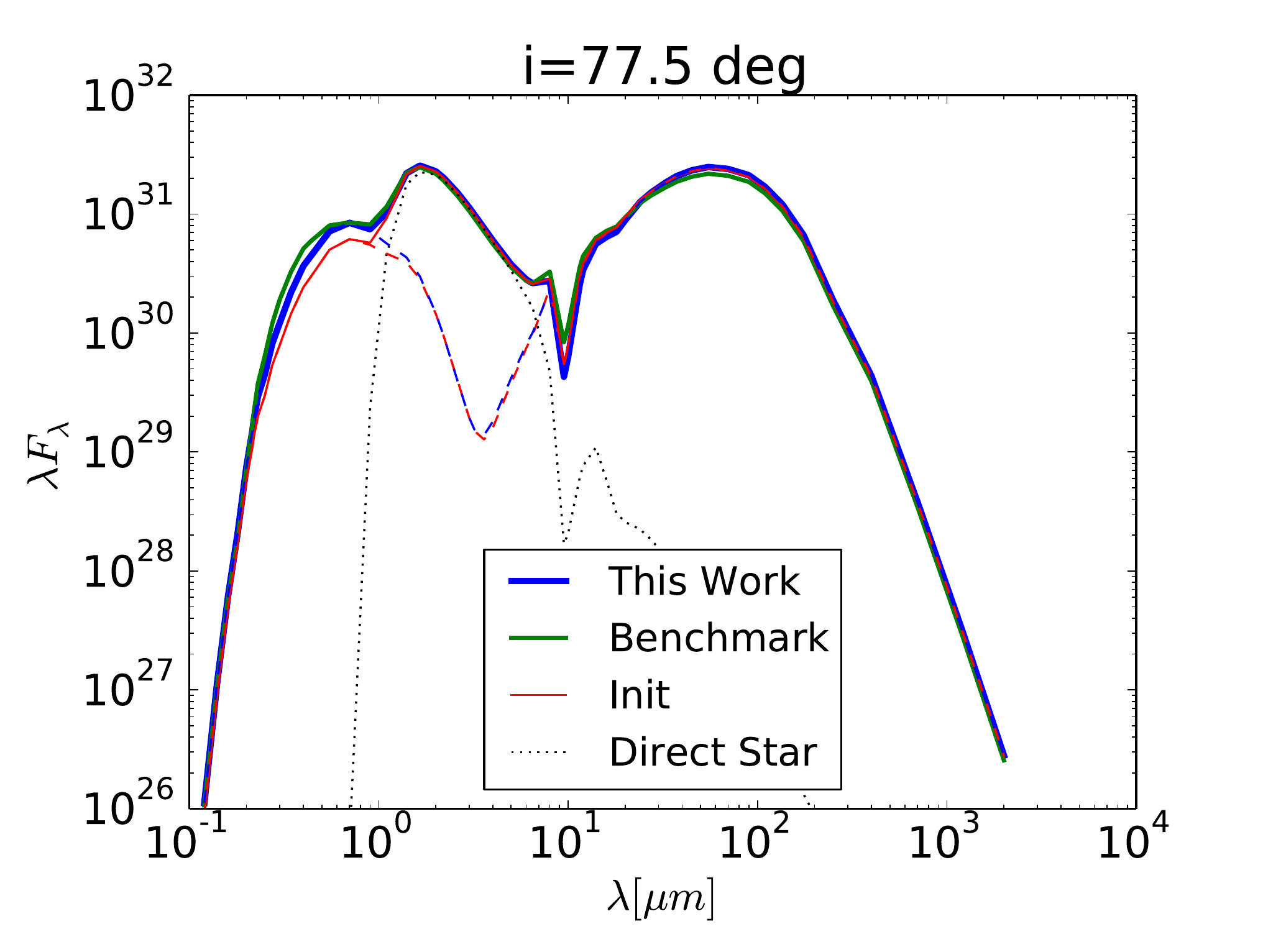}}

\caption{Comparison of HERO with the most difficult $\tau=100$ dusty torus benchmark test of \citealt{pascucci04}.  The top two panels show slices of the self-consistent temperature solution (top left: radial slices, top right: poloidal slices).  We also show the SC solution evaluated for two different choices of grid resolution: $N_\theta = (53, 106)$.  We find that the SC solution systematically underestimates the radiation/temperature field near the polar regions, with the effect becoming enhanced at high resolutions.  This is a consequence of the angular diffusion scheme used in SC (see Appendix \ref{app:diffusionscheme} for more discussion).  The bottom two panels compare spectra at different inclination angles as computed from LC raytracing.  
\label{fig:3D-Pascucci-SC}}
\end{center}  
\end{figure*}

Given the above boundary conditions, HERO solves for the radiation field everywhere, both inside and outside the disc, as well as the self consistent disc temperature, i.e. the temperature that satisfies Eq. \ref{eq:Tequilibrium}.  To benchmark the code, we consider the most difficult example presented in \citealt{pascucci04}, the case corresponding to $\tau = 100$.  The upper panels of Figure \ref{fig:3D-Pascucci-SC} compare the disc midplane temperatures computed by HERO with the benchmark models.  Overall, the agreement is reasonable -- the slight differences are likely due to ray defects.  We recover the temperature structure to within $10\%$, with the worst cells being located in the optically thick disc midplane.

The equilibrium temperature profile is highly sensitive to the amount of scattered light (which dominates over the direct stellar illumination near the disc surface by an order of magnitude). The good agreement between HERO and \citet{pascucci04} indicates that : 1) HERO correctly handles/redistributes the scattered light; 2) the temperature solver is robust.

As part of this test, we also show in Fig. \ref{fig:3D-Pascucci-SC} how the short and long characteristics version of HERO perform in determining the self-consistent disc temperature.  Short characteristics does a reasonable job everywhere within the optically thick parts of the disc, but systematically underestimates the radiation field and temperature in the optically thin regions.  Generally, SC has difficulties propagating radiation towards the coordinate poles.  Panel 2 of Figure \ref{fig:3D-Pascucci-SC} shows the resultant underestimated radiation temperatures ($\sim10$\% error) and that the bias increases with increasing $\theta$ resolution.

The long characteristics solver does not suffer from this systematic error and recovers the correct temperature profile within tens of iterations throughout the optically thin region.  This confirms that it is generally a good idea to run a few iterations of long characteristics after the short characteristics solver has been run to convergence.  We make a special point to emphasize the importance of the LC pass.  One might be tempted to instead run a high resolution SC pass to pin down a more accurate radiative solution. However we find that high resolution does not reduce the systematic biases that plague the SC method.  Despite the computational expense incurred by the LC method, it is the only way to obtain an accurate solution to the radiation field.  

Finally, we also show in the lower two panels of Fig. \ref{fig:3D-Pascucci-SC} integrated disc spectra.  Agreement between HERO and the other benchmark codes is within $20\%$, which is comparable to the spread amongst the four independent codes discussed in \citet{pascucci04}.  The HERO spectra were computed by raytracing from a distant observing plane, making use of the complete radiative solution (i.e. $S_\nu$) obtained from our SC+LC hybrid solver.

\subsection{GR Solutions}\label{sec:GRsolutions}

\subsubsection{Light Bending}

\begin{figure*}
\begin{center}  

\subfigure{\includegraphics[width=0.33\textwidth]{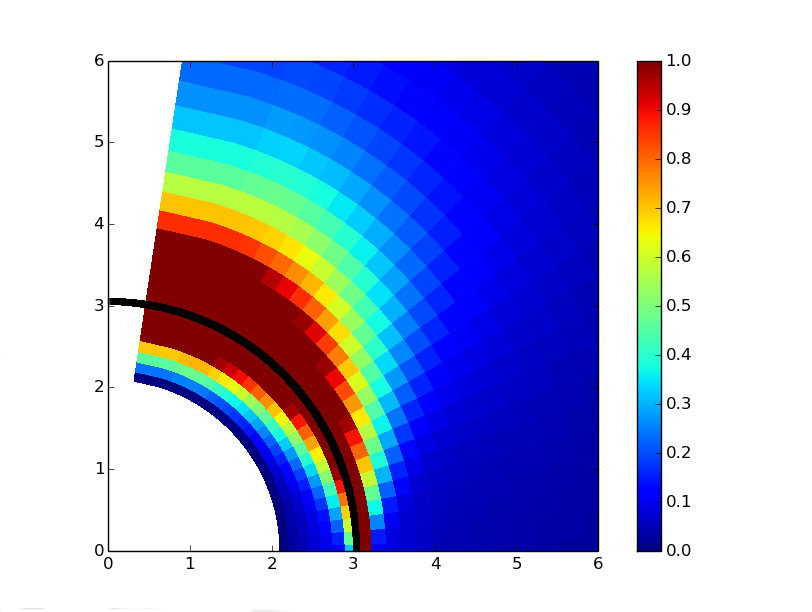}}
\subfigure{\includegraphics[width=0.33\textwidth]{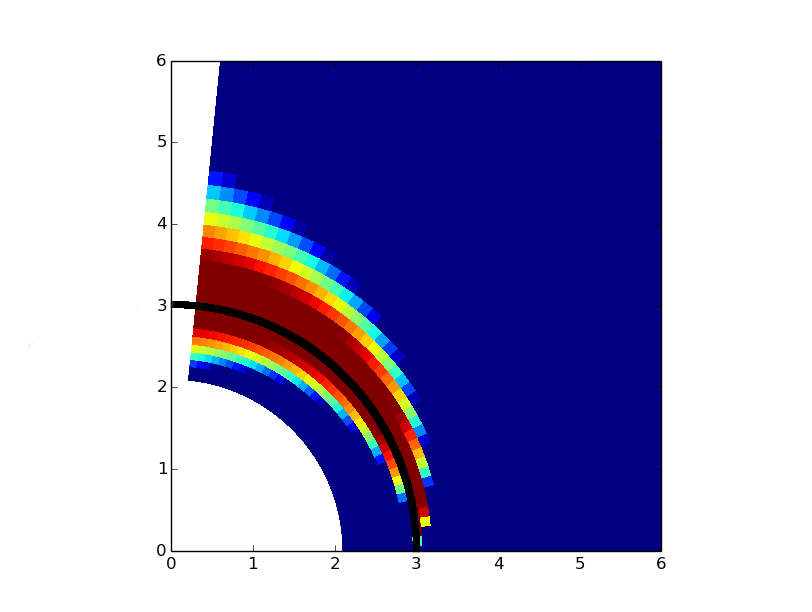}}
\subfigure{\includegraphics[width=0.33\textwidth]{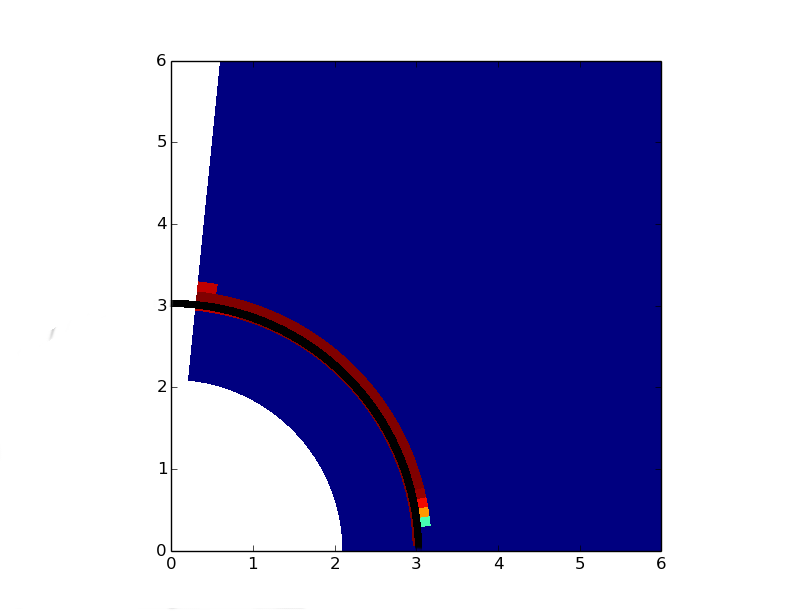}}

\caption{Light bending test for our two radiative solvers using a narrow laser beam injected tangent to the $r=3$ photon orbit.  From left to right: a) SC, b) LC with an artificially broadened beam to match the SC angle grid size, c) LC pure.  The solid black line shows the analytic result corresponding to the null geodesic.
\label{fig:3D-lightBend}}
\end{center}  
\end{figure*}

HERO is designed to solve for the radiation field in general relativistic curved spacetimes.  One important effect is light bending which is demonstrated in the test problem shown in Figure \ref{fig:3D-lightBend}.  $J$ is computed with both the SC and LC in the 3 panels for a beamed light source located just outside the photon orbit ($r=3M$) propagating in Schwarzschild space-time.  We see that the regions with the highest intensity follow the expected curved trajectory.  However, the beam is broadened substantially. This is a consequence of the finite angular grid (number of angles $N_A = 80$) used in the computation.

The middle and right panels of Figure \ref{fig:3D-lightBend} show LC solutions to the same problem.  The LC beam remains narrow and coherent, agreeing very well with the expected behaviour for free-streaming radiation at the photon orbit.  The middle panel shows the LC result using an emitting source whose beam has been artificially broadened to match the angular resolution of the SC $N_A=80$ angular grid.  The right panel shows the true resolving power of LC, where the emitting source corresponds to a $\delta$-function in angle.

\subsubsection{Disc Spectra}\label{sec:rayDefects}

To check that our handling of Doppler and gravitational redshifting is correct, we consider the problem of black hole accretion disc spectra.  This problem has been tackled numerous times over the years using many independent codes (e.g. \citealt{li05,davis06,kulkarni11,zhu12}) and is a simple but useful benchmark test.

We place an optically thick, geometrically thin disc around a Schwarzschild ($a=0$) black hole radiating as per the idealized \citet{NT} disc model. In this problem, the disk emission is treated as isotropic thermal radiation with flux given by the NT model.  For simplicity, we ignore any spectral hardening effects since we treat the emission in HERO as emanating from a single equatorial grid cell (i.e. for this test, we do not resolve the photon diffusion process that gives rise to spectral hardening).  The inner edge of the disk is fixed at the ISCO  $r_{\rm ISCO}=6$ and the outer edge is located at $r=1000$.

We solve for the radiation field in the disc exterior using HERO in full GR.  In HERO, the calculation of the disc spectrum is typically handled in two stages -- 1) we first solve for the 3D radiation field above the disc using the short/long characteristics solver (this is not needed in the present example because we specify the disc emission profile and assume vacuum outside the disc), and 2) we trace rays backwards (via Eq. \ref{eq:shortCharRT}) from a distant  observer plane located to create a synthetic image of the disc.  

The HERO code solves for the full three-dimensional source function and radiation field within the $2M < r < 1000M$ spatial domain of the grid. This information is then fed into a separate raytracing subroutine.  In the raytracing stage, parallel rays are shot towards the disk distributed according to a squeezed logarithmic polar grid (the same setup as \citealt{kulkarni11}) from an observer plane located at $r=100,000M$.  The final spectrum is generated by integrating the flux across the observer plane.

\begin{figure}
\begin{center}  
\subfigure{\includegraphics[width=0.5\textwidth]{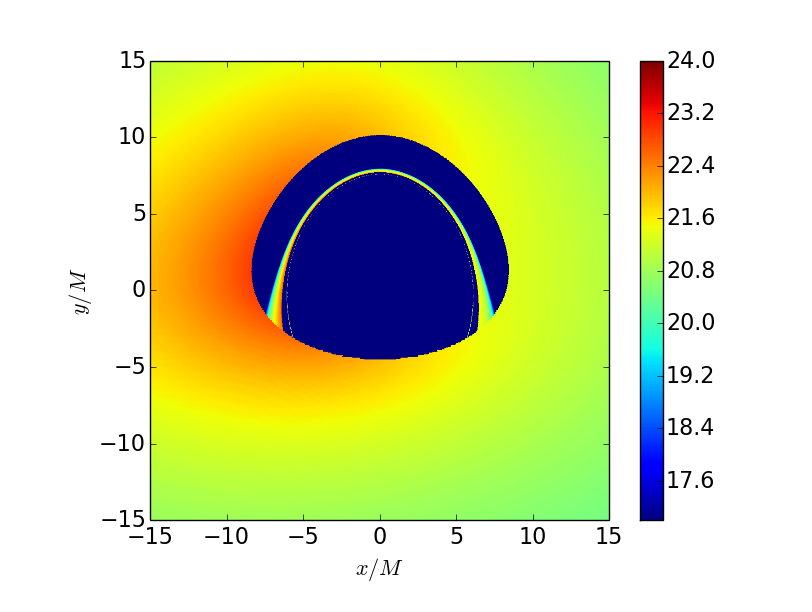}}
\subfigure{\includegraphics[width=0.45\textwidth]{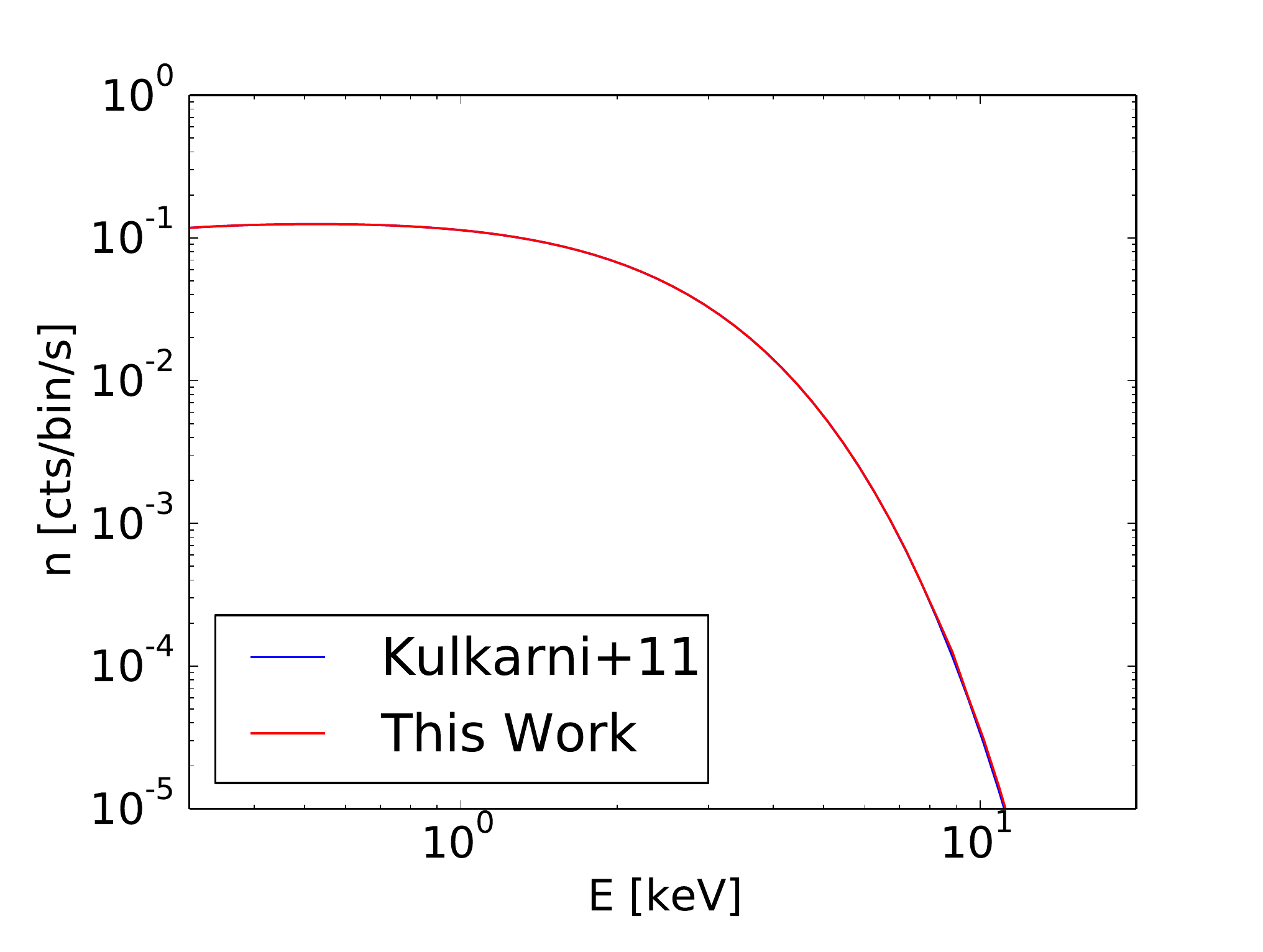}}
\subfigure{\includegraphics[width=0.45\textwidth]{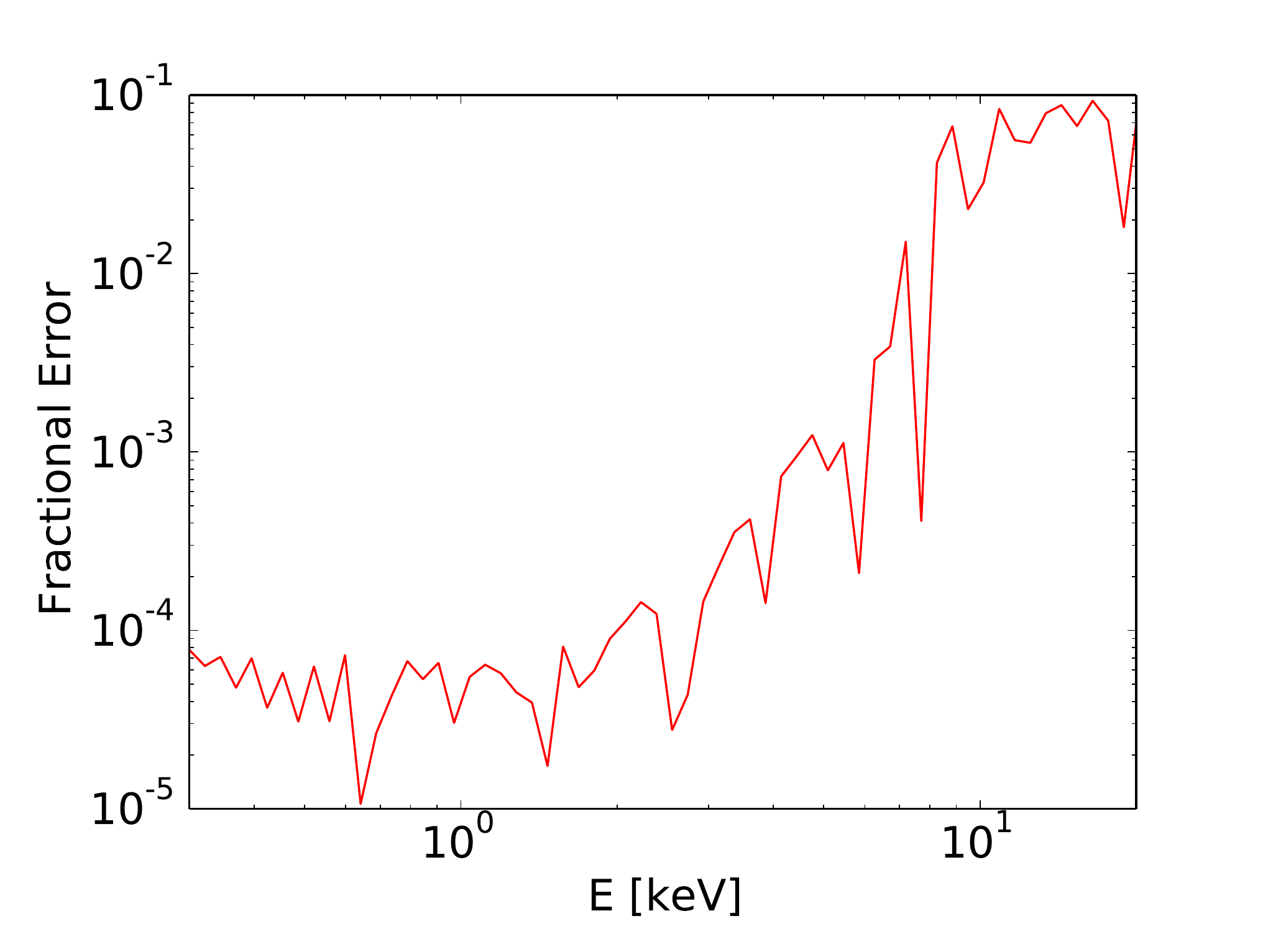}}
\caption{Upper panel: Raytraced image of a razor thin disc as viewed by an observer located with inclination angle $i=60$.  Colors indicate log of the frequency integrated intensity.  The asymmetry is due to the doppler effect combined with gravitational redshifting.  The central arc feature corresponds to a secondary image of the accretion disk that arises from strong gravitational lensing about the black hole.  Middle panel: Integrated disc spectrum as computed by HERO (red points), compared to the result with the \citealt{kulkarni11} code (blue line).  Bottom panel: fractional errors between HERO and \citealt{kulkarni11}.
\label{fig:3D-spectra}}
\end{center}  
\end{figure}

The top panel of Figure \ref{fig:3D-spectra}) shows the computed image for the particular example problem.  Integrating over this image yields the observed disc spectrum, which we show in the lower panel.  The disc spectrum computed with HERO agrees very well with a previous calculation by \citet{kulkarni11}, lending confidence that HERO correctly handles GR effects.

\subsubsection{Iron Line Spectra}

In a similar vein as for the previous test, which was based on the NT continuum spectrum of the disk, we also benchmark our code via a calculation of the emission profile due to a monoenergetic line (e.g. Fe-K$\alpha$ as seen in many Seyfert galaxies and some microquasars).  For a system that emits only monoenergetic $\delta$-function lines, the final integrated line profile depends only on the geometrical properties of the system (i.e. redshifting from doppler/gravitational curvature and lensing effects from the Kerr metric).

\begin{figure}
\begin{center}  
\subfigure{\includegraphics[width=0.45\textwidth]{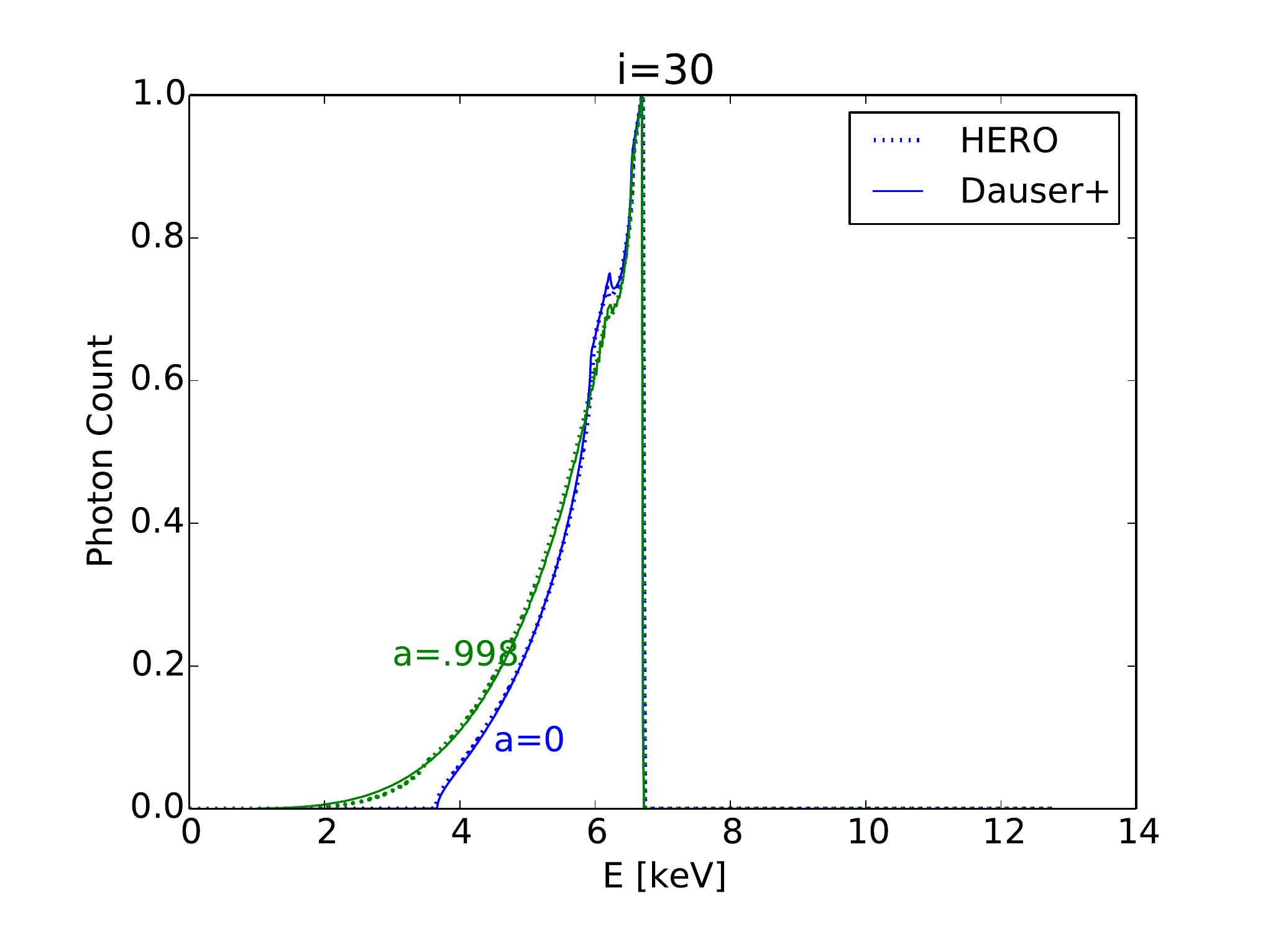}}
\subfigure{\includegraphics[width=0.45\textwidth]{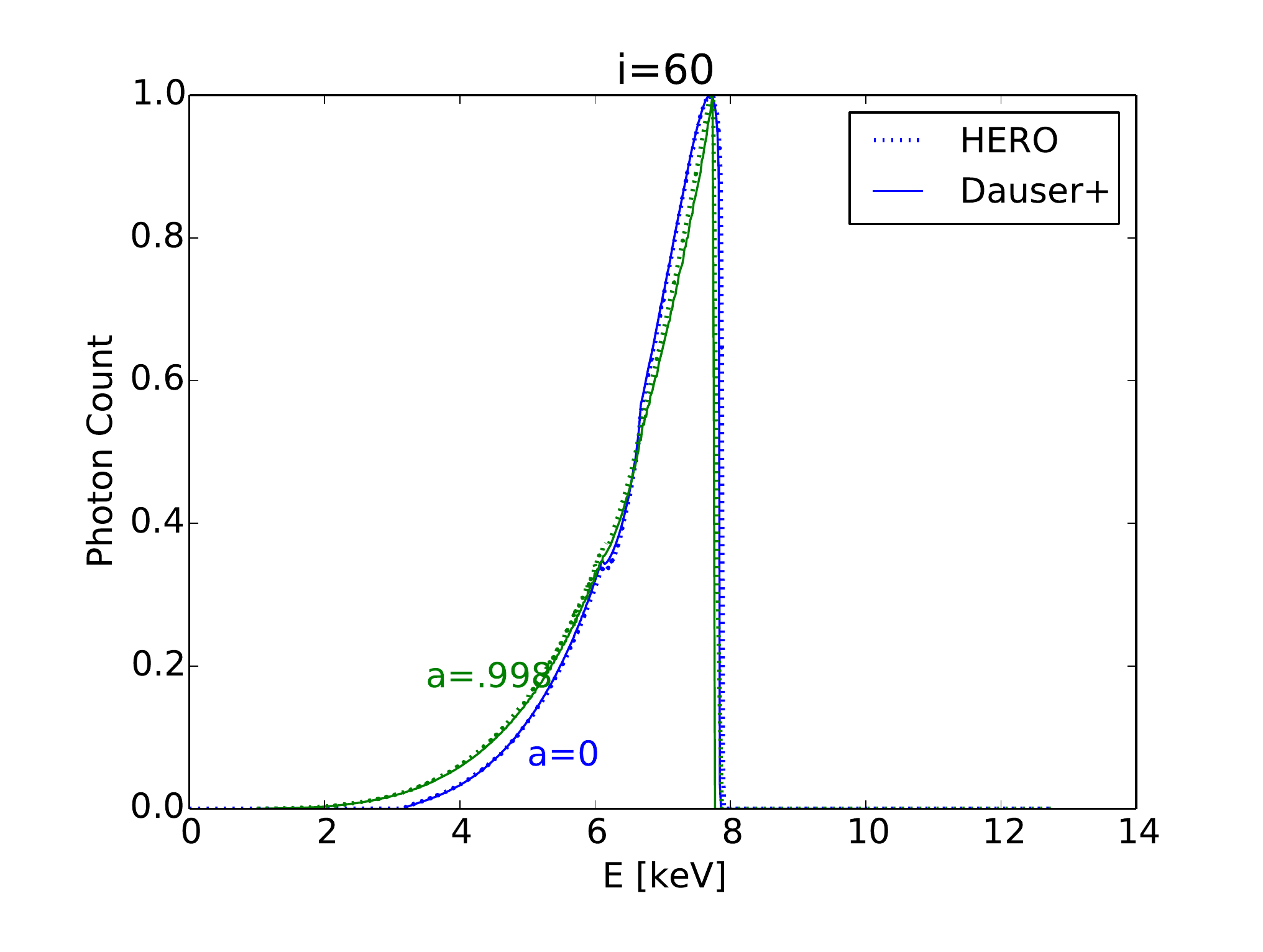}}
\caption{A comparison of the Fe-K$\alpha$ line profile as computed by HERO and RELLINE \citep{dauser13} for nonspinning (blue, $a/M=0$) and nearly maximally spinning (green, $a/M=0.998$) black holes.  Line profiles are computed for different viewing inclinations: top panel shows $i=30$, whereas bottom panel shows $i=60$.
\label{fig:iron-line}}
\end{center}  
\end{figure}

We consider a Keplerian accretion disk in the Kerr metric whose line emission is modulated by a power law emissivity profile $F(r) \propto r^{-3}$.  We also set the domain of the disk to span from $r_{in} = r_{ISCO}$ out to $r_{out} = 400$.  Figure \ref{fig:iron-line} shows the resultant spectra computed for two different choices of BH spin and two different choices of viewing angle.  We find good agreement between HERO and the benchmark code RELLINE \citep{dauser10, dauser13}, with the largest discrepancies occurring in the low energy red tail of the line profiles.

The setup in HERO is identical to that of the previous NT disc tests, except that a monoenergetic 6.4\,keV line was used as the local emissivity instead of a thermal continuum source.  In addition, we use much higher resolution for the raytracing grid since the low energy red tail of the line profile depends sensitively on how well the inner edge of the disc is resolved.

\subsubsection{Returning Radiation}
Another classic accretion disk problem is that of computing the returning radiation due to relativistic light bending around the black hole.  For this test, we setup a razor thin accretion disk that emits according to the standard thin disc luminosity profile \citep{pagethorne74} and measure the amount of returning radiation incident on the disk.  The goal is to benchmark both our SC and LC radiative solvers against the solutions of \citet{cunningham76}, who tackled the same problem by means of relativistic transfer functions.

In HERO, we model the returning radiation problem with a grey calculation (1 bolometric frequency bin) on an axisymmetric polar grid with ($n_r,n_\theta$)=(60,30) restricted to the upper half plane.  The spatial grid was set as uniformly spaced in angle and $\log(r)$.  Boundary conditions invoked are: reflecting for the polar axis (to account for light that passes through the pole), constant flux injection at the equatorial disk plane according to \citep{pagethorne74}, and zero incident radiation at the inner and outer radial boundaries.

\begin{figure}
\begin{center}  
\subfigure{\includegraphics[width=0.45\textwidth]{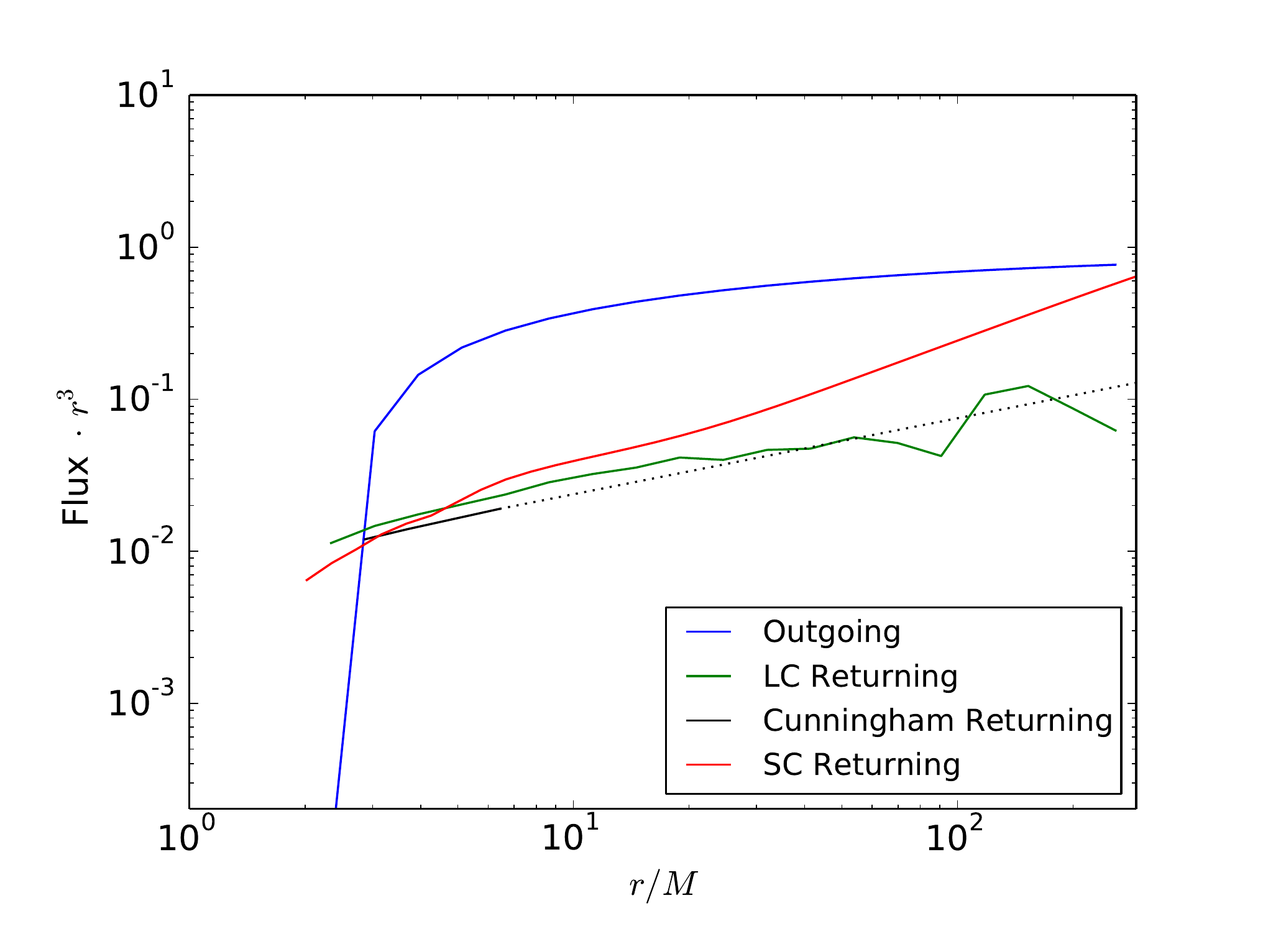}}
\caption{A calculation of the returning radiation for a razor thin accretion disc around a moderately spinning $a_*=0.9$ black hole.  The standard \citet{pagethorne74} luminosity profile is used to set the outgoing flux/rays from the disk plane (blue).  We compare the returning flux as calculated by the two radiative solvers in HERO (red, green) with the result from \citet{cunningham76} obtained via relativistic transfer functions (black).  The black dashed line is an extrapolation of the values from Cunningham's data table.
\label{fig:returningRadiation}}
\end{center}  
\end{figure}

In figure \ref{fig:returningRadiation}, we plot the incoming radiation flux at various locations above the disk and compare to the solution of \citet{cunningham76} for a moderately spinning $a_*=0.9$ black hole.  We find that SC systematically overestimates the amount of returning flux at large radii, presumably caused by the angular interpolation bleeding some of the outbound radiation into inbound rays.  LC on the other does not experience any strong systematic biases, but suffers from a lack of resolving power at large radii since at these large distances, the LC ray grid can miss the inner photon ring that is responsible for most of the returning radiation.

\subsubsection{Vacuum Test}\label{sec:vacuumtest}

\begin{figure}
\begin{center}  
\subfigure{\includegraphics[width=0.45\textwidth]{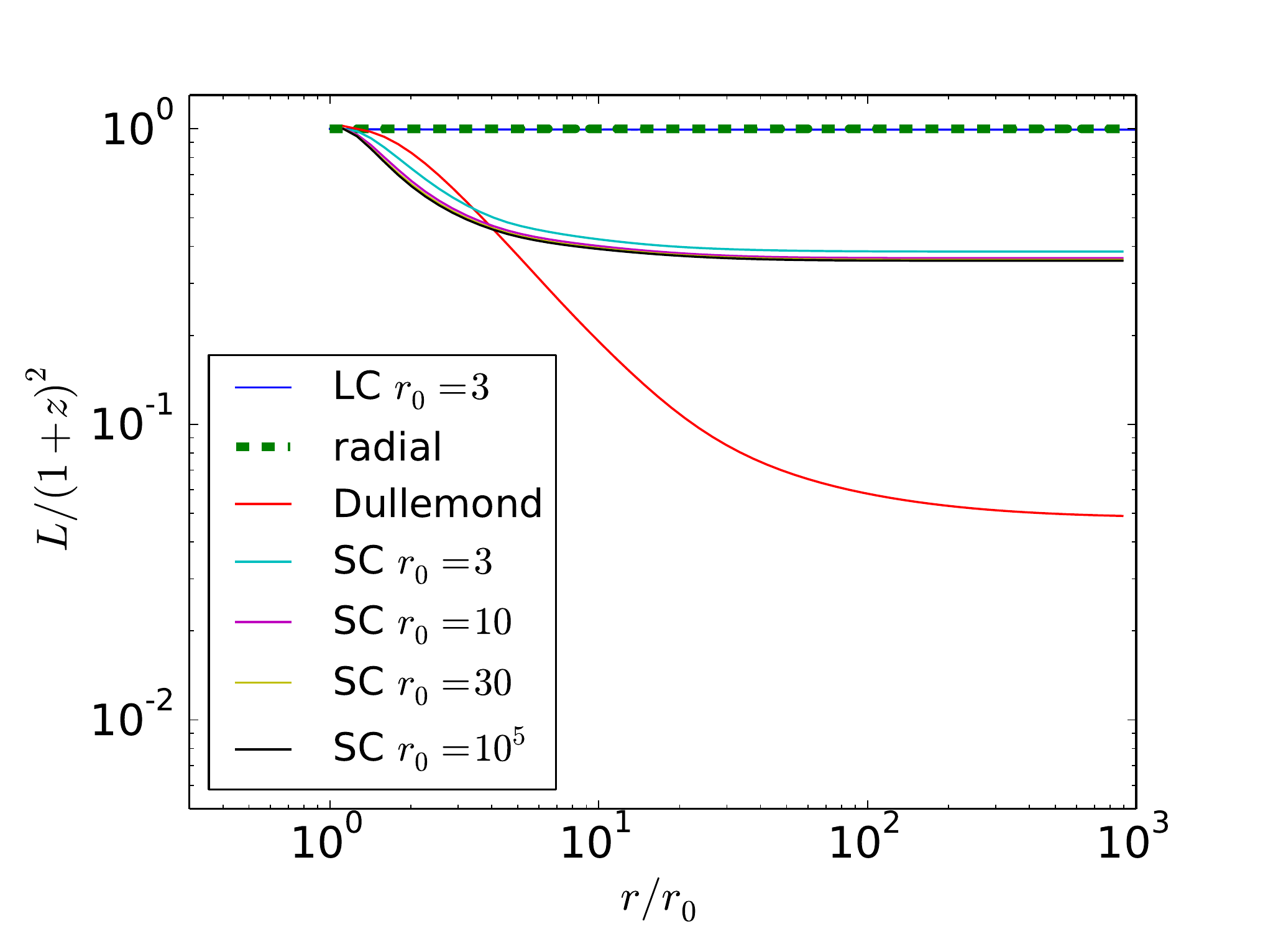}}
\subfigure{\includegraphics[width=0.45\textwidth]{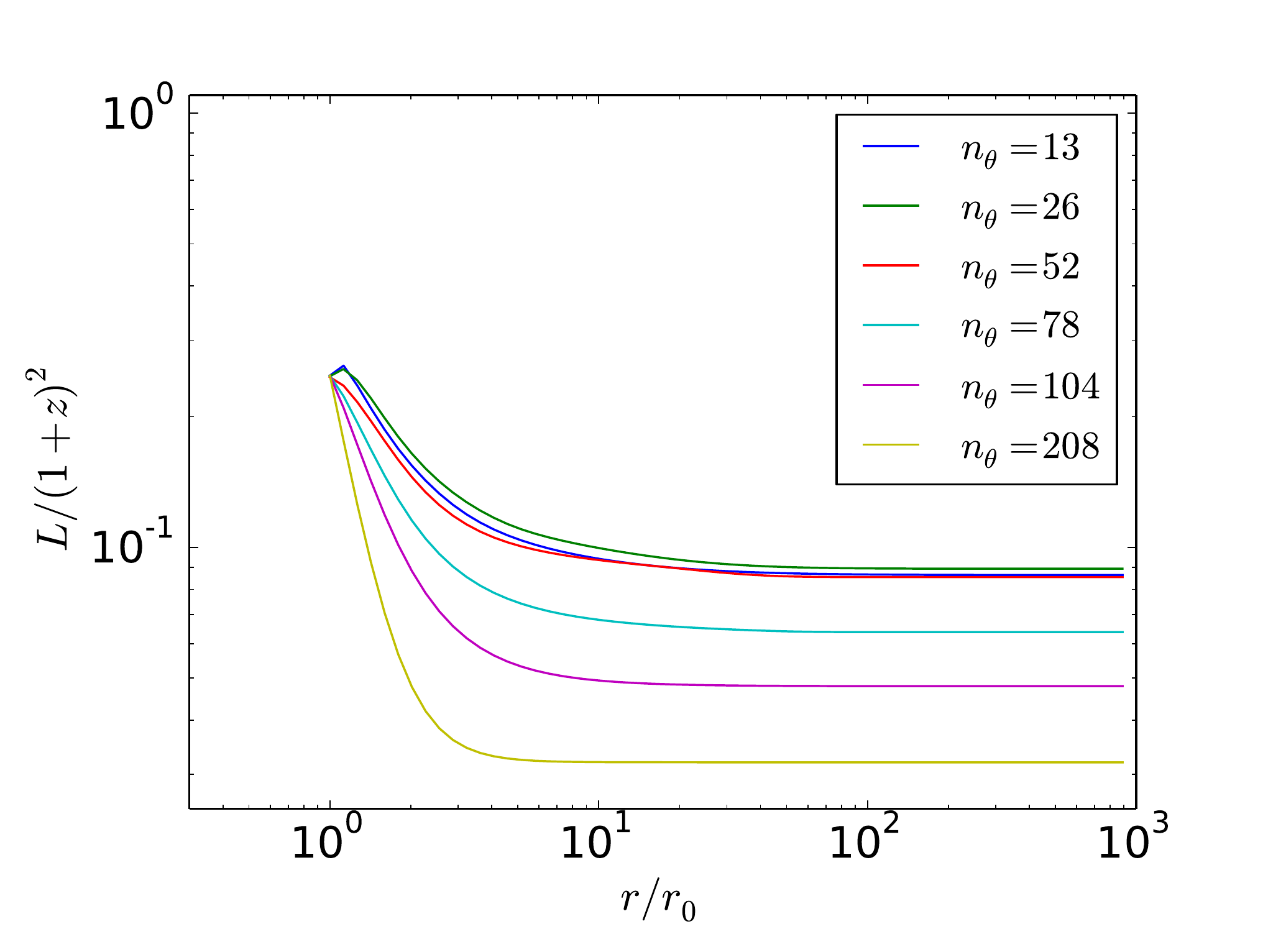}}

\caption{Variation in radial luminosity profiles as a function of a) relativistic effects (lensing, redshifting) and b) grid-$\theta$ resolution.  Note that the luminosity plotted is the corrected luminosity as measured by an observer at infinity.  The analytic solution is therefore a constant $L/(1+z)^2$ for all radii. 
\label{fig:vacuumtest} }
\end{center}  
\end{figure}

As a final test of the general relativistic capabilities of HERO, we turn to the problem of light propagation in vacuum from an opaque spherical shell.  In Schwarzchild geometry, it is a simple exercise to solve for the apparent angular size of a constant radius shell as viewed by an observer at some other (larger) radius.  Furthermore, if the surface of the shell radiates isotropically like a blackbody with a constant surface temperature, one can calculate the radiation quantities (i.e. $J, F, L$) by simply integrating the constant intensity across the apparent solid angle subtended by the shell.   For instance, the radial profile of luminosity has a particularly simple form, scaling directly with gravitational redshift as
\begin{align}
L_{\rm{local}} &= L_\infty (1+z)^2
\end{align}
where $L_{\rm{local}} = F_{\rm{local}} 4 \pi r^2$ is the total luminosity as measured by an observer at radius $r$ and $L_\infty$ is the luminosity at infinity.  In Figure \ref{fig:vacuumtest}, we show the results for the luminosity as calculated by our general-relativistic short and long characteristics solvers.  We normalize the luminosity by the redshift factors so the analytic solution simply corresponds to a flat horizontal line (i.e., constant luminosity as measured at infinity).  

In general, we find that our short characteristics solver systematically underestimates the luminosity profile at large distances.  These tests were carried out with the diffusion prescription described earlier, therefore if the shell radiated purely in the radial direction, SC would by construction give the correct answer. Here the surface radiates isotropically, and there is a deviation from the true answer because of angle interpolation and diffusion.  The various SC curves in Figure \ref{fig:vacuumtest} correspond to different choices for the radius of the inner emitting shell.  The high degree of similarity for all choices of inner radius (i.e. ranging from from highly relativistic $r_0=3$ in units of $GM/c^2$ to nonrelativistic $r_0 = 10^5$) implies that the SC bias is independent of relativistic effects.  It arises purely from the discretization of the spatial and angular grid. While the bias is not negligible -- the luminosity is reduced by a factor of 2.5 at large radius -- note that the luminosity would be a factor $\sim10^6$ too large if we did not include diffusion. As with the other test problems, LC gives an essentially perfect answer.

Panel two of Figure \ref{fig:vacuumtest} shows another interesting phenomenon, viz., there is a strong dependence between the SC bias on the choice of spatial grid resolution.  As we increase $\theta$-resolution (keeping the $r$-resolution fixed), the luminosity profile of our isotropically emitting shell exhibits a stronger bias to lower values.  This effect is ultimately caused by the angle diffusion scheme that we employ (c.f. Appendix \ref{app:diffusionscheme}).  The diffusion is tuned such that the radial cell-to-cell attenuation of light recovers the inverse-square law.  If the cell aspect ratio in $r-\theta$ is too rectangular, then the diffusion has a strong directional preference.  Light rays traveling along the short dimension of the cell are overattenuated since they hit cell boundaries more often than rays traveling in the long-dimension (the degree of diffusion is directly proportional to how often the light ray traverses cells within a given spatial distance).  Based on our tests,  grid cells should ideally have a square aspect ratio to minimize the error, and any ratio in excess of 2:1 should be avoided.

\section{Summary}\label{sec:conclusion}

We have described in this paper HERO, a new general relativistic radiative transfer code.  The primary aim of this code is to model the radiation field in accretion flows around black holes.  The unique features of HERO are: 1) a hybrid short/long characteristics radiative solver that enables accurate and fast modelling of complex anisotropic radiation fields; 2) implementation in a general relativistic framework taking into account the effects of light bending, doppler beaming, and gravitational redshifting.

HERO is written as a post-processing code decoupled from the hydrodynamic evolution of the fluid.  It computes the time-independent radiation field assuming a given fixed background fluid structure.  Strictly speaking, this approach is valid only for problems where the fluid velocities are small compared to the speed of light (i.e. nonrelativistic flows).   Alternatively, and this is the primary application we have in mind, it could also be applied to time-steady relativistic problems.

We provide a detailed explanation of the long/short characteristics method used to solve for the radiation field and our approach for solving the self-consistent gas temperature.  To verify that HERO produces physically correct answers, we have performed a comprehensive set of tests designed to examine the code's convergence properties, accuracy, and capability to handle multidimensional relativistic problems.  We confirm the well known result that 2D and 3D problems with compact sources suffer from significant ray defects in the far field when analyzed with the short characteristics method.  We present an approximate fix which mitigates the effects in the case of a 3D spherical grid.  However, for accurate results, it is necessary to switch to a long characteristics solver which is unaffected by ray-defects.

As the subject of a follow-up paper, we intend to apply HERO to radiative MHD simulations of accretion discs -- particularly simulations undergoing super-Eddington accretion, where radiation feedback strongly dominates the dynamics of the flow.  Using HERO, we will investigate the integrated spectra to see the role that self-shadowing and irradiation plays in these systems.  This application requires a Comptonization module which will be described in our next follow-up paper.

\section{Acknowledgements}

The authors would like to thank Nathan Roth, Jack Steiner, Jonathan McKinney, James Guillonchon, Jeff McClintock, Yan-Fei Jiang, Javier Garcia, Eric Keto, and Jiachen Jiang for their excellent insights on radiative transfer and comments/suggestions regarding HERO.  We also thank our anonymous referee for their many suggestions that have greatly improved the cohesiveness of this paper (especially their recommendations of several relativistic benchmark tests).  RN and YZ akcnowledge support from NSF grant AST1312651 and NASA grant NNX 14AB47G.  AS acknowledges support for this work by NASA through Einstein Post-doctoral Fellowships PF4-150126, awarded by the Chandra X-ray Center, which is operated by the Smithsonian Astrophysical Observatory for NASA under contract NAS8-03060.  Finally, we are grateful for support from NSF XSEDE grant TG-AST080026N, the NASA HEC Program, and the Harvard Odyssey cluster for providing computing resources used in developing this code.\\


\bibliography{ms}

\appendix

\section{Ray Defects}\label{app:raydefects}

\begin{figure}
\begin{center}  
\subfigure{\includegraphics[width=0.45\textwidth]{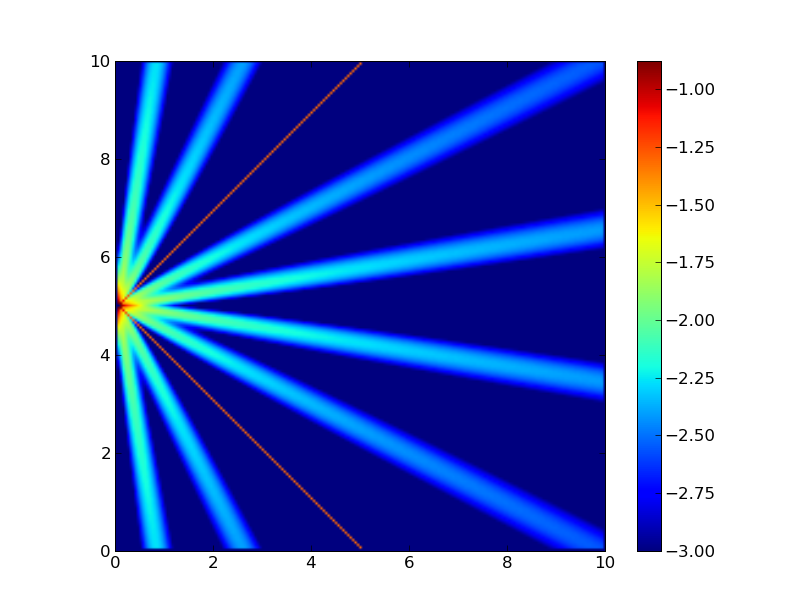}}
\subfigure{\includegraphics[width=0.45\textwidth]{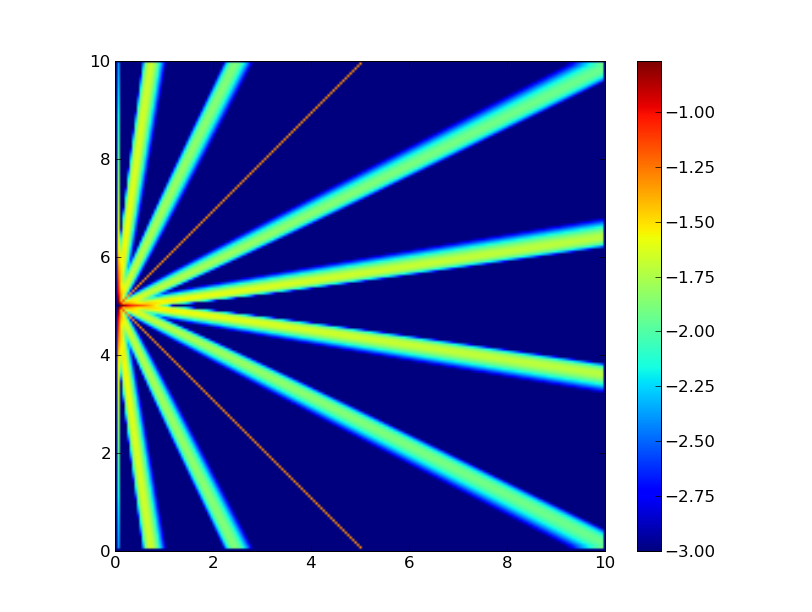}}
\subfigure{\includegraphics[width=0.45\textwidth]{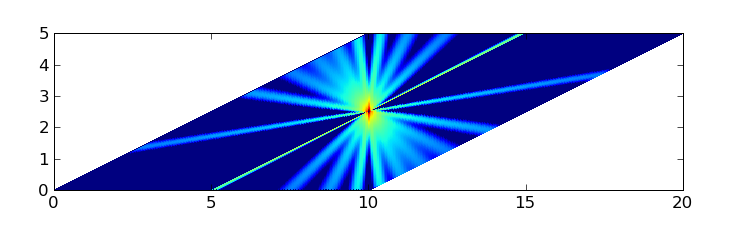}}
\subfigure{\includegraphics[width=0.45\textwidth]{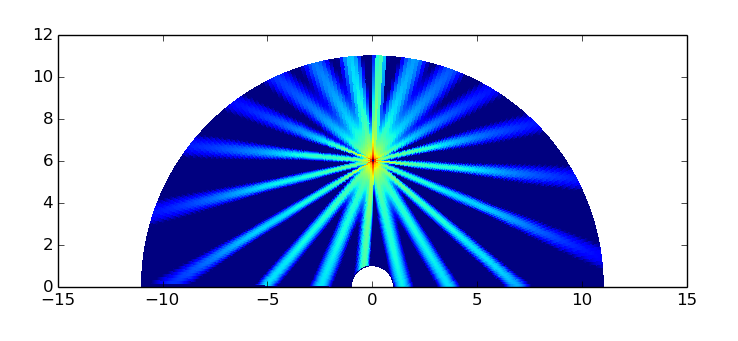}}
\caption{Comparison of the radiation solution for a point source computed by short characteristics for different grid choices ($N_A=20$). Colour indicates $\log(J)$.  From top to bottom: a) cartesian grid (100x100); b) same as a), but using quadratic interpolation; c) sheared cartesian (100x100); d) Polar grid (60x100).  In all cases, we place a delta function source function that emits isotropically.  The pencil beams that appear are the ray-defects -- physically, the far-field radiation pattern should be spherical and smooth but the limitations of treating radiation only locally in the SC method results in the thin beams (see discussion in \S\ref{app:defectOrigin}).
\label{fig:ray-defects}}
\end{center}  
\end{figure}

The method of short characteristics constitutes the primary workhorse of our radiative solver and is used to generate a good first approximation to the radiation field.  However, ``ray-defects" are a well known limitation of SC, which is why we need to follow up SC with the more accurate LC method. Here we discuss some of the properties and explore the cause of ray defects.  Figure \ref{fig:ray-defects} shows a few examples of ray defects in 2D arising from point source emitters.  The defects manifest as unphysical beam-like patterns far from the emitting source.

Point sources generate the strongest defect pattern so we use them in the following discussion to illustrate the main issues.  Point-source-like emission does appear in the problem of black-hole accretion discs, e.g. the hottest innermost region of the disc shines extremely brightly and due to its compact spatial scale, acts like a point source at large distances from the centre. 

For a fixed angular grid, the point source ray defects form a series of radial beams that reflect the underlying structure of the angular grid (see Figures \ref{fig:ray-defects} for a few examples in different coordinate systems). Beam collimation is enhanced for rays travelling in directions where neighbouring grid cells cover a smaller angular size.  This "grid-lattice" effect is best seen in panel c) (sheared box) in Figure \ref{fig:ray-defects}, where the thinnest beams are those travelling to the upper-right/lower-left sectors (i.e. the directions where the neighbour points as defined in Figure \ref{fig:ShortCharSchematic} span the smallest angular extent).  The same, but less pronounced result is seen in panel d) of Figure \ref{fig:ray-defects} (polar coordinate system).  Here, the rays pointed radially inward suffer less dispersion than their radially outward counterparts, again for the same reason as in the sheared box case (tighter angular packing occurs for the neighbour cells at smaller radius).

Finally, ray defects are particularly enhanced for rays that directly intersect neighbouring cell centres (as an example, note that the $45^o$ rays exhibit overwhelmingly strong defect patterns in our cartesian setup in the top two panels of Fig \ref{fig:ray-defects}).

Ideally, we would like to represent the discretized radiation pattern as a smooth field instead of a superposition of laser beams.  After much trial and error, we have arrived at an angular diffusion based solution for spherical log polar grids.  To motivate our final solution, we first examine the root cause of the defects.

\subsection{Mathematical Origin of Ray Defects}\label{app:defectOrigin}

Ray defects ultimately arise due to the local nature of the short characteristics solver.  Methods that operate using just local propagation of light develop a ``Pascal's Triangle" characteristic beam pattern with increasing distance.  If the radiation from a source cell is split with some linear combination to its neighbouring cells and the same method is applied uniformly for all cells, then the propagation of light has the following characteristic shape -- consider the case where the mixing coefficients are [1/2, 1/2]:

\begin{centering}

  \begin{tikzpicture}
    \def\numrows{5}
    \def\dx{20pt}
    \def\dy{30pt}
    \newcounter{i}
    \stepcounter{i}
    \node (\arabic{i}) at (0,0) {1};
    \foreach [count=\i] \x in {2,...,\numrows}{
      \pgfmathsetmacro{\lox}{\x-1}%
      \pgfmathsetmacro{\loxt}{\x-3}%
      \foreach [count=\j] \xx in {-\lox,-\loxt,...,\lox}{
        \pgfmathsetmacro{\jj}{\j-1}%
        \stepcounter{i}
        \pgfmathsetmacro{\lbl}{\lox!/(\jj!*(\lox-\jj)!)}
        \node  (\arabic{i}) at (\xx*\dx, -\lox*\dy) {\pgfmathint{\lbl}\pgfmathresult/$2^\i$};
      }
    }
    \newcounter{z}
    \newcounter{xn}
    \newcounter{xnn}
    \pgfmathsetmacro{\maxx}{\numrows - 1}
    \foreach \x in {1,...,\maxx}{
      \foreach \xx in {1,...,\x}{
        \stepcounter{z}
        \setcounter{xn}{\arabic{z}}
        \addtocounter{xn}{\x}
        \setcounter{xnn}{\arabic{xn}}
        \stepcounter{xnn}
          \draw [->] (\arabic{z}) -- (\arabic{xn});
          \draw [->] (\arabic{z}) -- (\arabic{xnn});
      }
    }
  \end{tikzpicture}
\begin{equation}
\vdots \nonumber
\end{equation}

\end{centering}
The final far field pattern simply corresponds to the weights of a discrete random walk.  This shape is a Gaussian, with a characteristic width of $w \propto \sqrt{d}$ where $d$ is the total propagation distance from the source.  The beam far from the source will therefore have angular size $w/d \propto 1/\sqrt{d} \rightarrow 0$ as $d\rightarrow \infty$.  Thus, at a sufficiently large distance, the propagation of light via short characteristics will inevitably result in a series of laser beams.

The defect problem can be further exacerbated if a ray happens to pass exactly through a neighbouring cell centre.  In this case, the mixing/interpolation coefficients are [1, 0] and the propagation diagram reduces to:

\begin{centering}

  \begin{tikzpicture}
    \def\numrows{3}
    \def\dx{20pt}
    \def\dy{30pt}
    \setcounter{i}{1}
    \stepcounter{i}
    \node (\arabic{i}) at (0,0) {1};
    \foreach [count=\i] \x in {2,...,\numrows}{
      \pgfmathsetmacro{\lox}{\x-1}%
      \pgfmathsetmacro{\loxt}{\x-3}%
      \foreach [count=\j] \xx in {-\lox,-\loxt,...,\lox}{
        \pgfmathsetmacro{\jj}{\j-1}%
        \stepcounter{i}
        \pgfmathsetmacro{\lbl}{\lox!/(\jj!*(\lox-\jj)!)}
          \ifnum\j<2
            \node  (\arabic{i}) at (\xx*\dx, -\lox*\dy) {1};
          \else
            \node  (\arabic{i}) at (\xx*\dx, -\lox*\dy) {0};
          \fi
      }
    }
    \setcounter{z}{1}
    \setcounter{xn}{1}
    \setcounter{xnn}{1}
    \pgfmathsetmacro{\maxx}{\numrows - 1}
    \foreach \x in {1,...,\maxx}{
      \foreach \xx in {1,...,\x}{
        \stepcounter{z}
        \setcounter{xn}{\arabic{z}}
        \addtocounter{xn}{\x}
        \setcounter{xnn}{\arabic{xn}}
        \stepcounter{xnn}
          \draw [->] (\arabic{z}) -- (\arabic{xn});
          \draw [->] (\arabic{z}) -- (\arabic{xnn});
      }
    }
  \end{tikzpicture}
\begin{equation}
\vdots \nonumber
\end{equation}

\end{centering}

Notice that for this limiting case the beam pattern does not spread at all, resulting in a zero width beam of constant intensity at all distances.  This kind of defect is particularly disastrous in the case of spherical coordinates since it affects all radial rays.  Thus, any compact light source near the centre of the spherical grid will develop serious ray-defects at larger radii. The effect is quite devastating if there is a strong point source at the centre.  The radial ray defect acts to force a constant intensity on all radial rays, resulting in constant radiation energy density independent of distance rather than the expected inverse square falloff.  The natural way to correct non-spreading beams is to introduce some degree of artificial broadening (i.e. force the mixing coefficients to have some minimum floor value).  We describe our approach in detail in the next section.

\subsection{Ray Defect Correction Schemes}\label{app:diffusionscheme}

The most obvious approach to mitigate the impact of ray defects is simply increasing the angular resolution.  However this approach is not feasible for complex problems with large spatial domains due to the unfavorable computational cost scaling with resolution.

One idea to combat ray defects is to use higher order interpolation schemes in treating the propagation of radiation.  \citet{davis12} found that quadratic monotonic interpolation works well and suggest that it be used over standard linear interpolation.  Unfortunately, this does not address the fundamental problem of beam spreading -- higher order schemes are actually counterproductive in that they produce even thinner beam patterns and amplify the significance of ray defects (Compare the top two panels of Figure \ref{fig:ray-defects} for linear vs. quadratic interpolation) .  

Ultimately, the problem (as illustrated in \S\ref{app:defectOrigin}) has to do with the lack of spreading for a single beam.  This motivates us to implement a diffusive scheme to bring back the necessary amount of spreading and to set the intensity values in such a way that an inverse square falloff of radiation density and flux is recovered.

Our approach is the following: every time we read out intensity values from our angular grid (i.e. the interpolation scheme described in \S\ref{sec:angleInterp}), we apply a floor to our interpolation coefficients so that there is always some degree of mixing between nearby angles.  That is, we modify the coefficients $c_1, c_2, c_3$ in Eq. \ref{eq:angleInterp} to new values $c_1', c_2', c_3'$ given by:
\begin{equation}
c_i' = \frac{c_i + w}{\left(\sum\limits_{i}c_i\right) + 3w} \label{eq:angleDiffusion}
\end{equation}
where $w$ is the diffusion coefficient that controls the amount of angle mixing.  A more physical way to understand this diffusion coefficient is to convert it into an attenuation factor for nearby cells.  For a monodirectional beam (i.e. $I=0$ for all but a single angle bin), the beam will attenuate by a factor $a$ after propagating across a single cell, i.e.,
\begin{equation}
I^{n+1} = a I^n,
\end{equation}
where from Eq. \ref{eq:angleDiffusion}, we find:
\begin{equation}
a = (1+w)/(1+3w).
\end{equation}
This motivates a method for choosing the appropriate value of $w$ for the case of a log-polar grid (where the radial spacing is uniform in log$r$).  Since we desire an inverse square law falloff of the radiation field, we simply solve for $a$ in the equation $a^{N_{dec}} = 0.01$ where $N_{dec}$ is the number of cells per radial decade.  For example if $N_{dec} = 20$ points per decade (our canonical choice), we have
have $a = 0.794$ and $w=0.149$.

\begin{figure}
\begin{center}  
\subfigure{\includegraphics[width=0.4\textwidth]{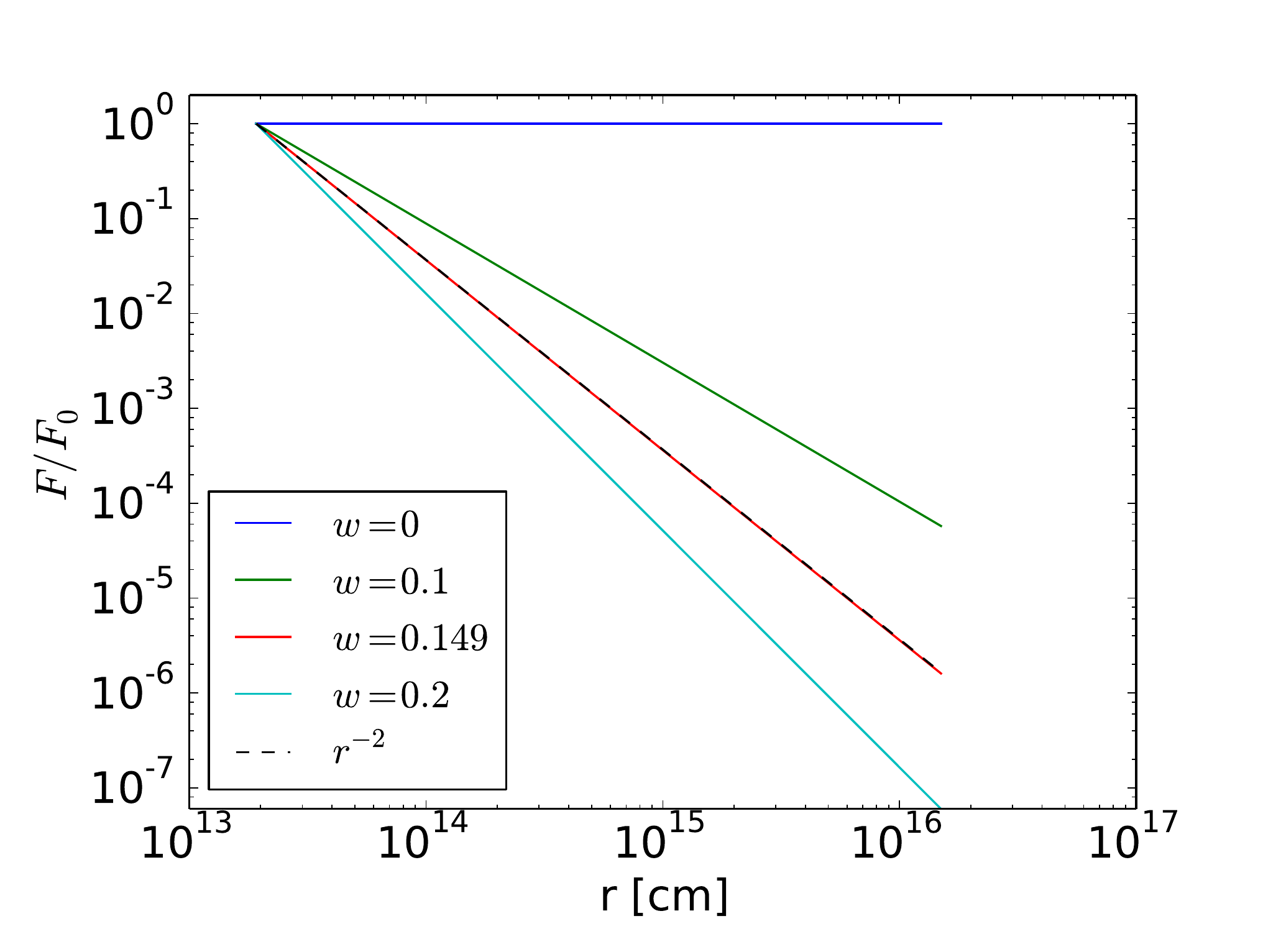}}
\caption{Radial profiles of $F$ for various choices of diffusion coefficient as calculated by short characteristics code on a spherical polar grid $(n_r,n_\theta)=(60,100)$.  Given these grid dimensions, the diffusion coefficient must be set to $w=0.149$ in order to reproduce an inverse square falloff for flux.
\label{fig:radial-falloff}}
\end{center}  
\end{figure}

In Figure \ref{fig:radial-falloff}, we show radial profiles of $F$
calculated with the SC for a point source central star.  Note that, in the absence of diffusion,
the effect of radial ray defects is to produce an unphysical solution
where $F$ is constant with radius.

\begin{figure*}
\begin{center}  
\subfigure{\includegraphics[width=0.3\textwidth]{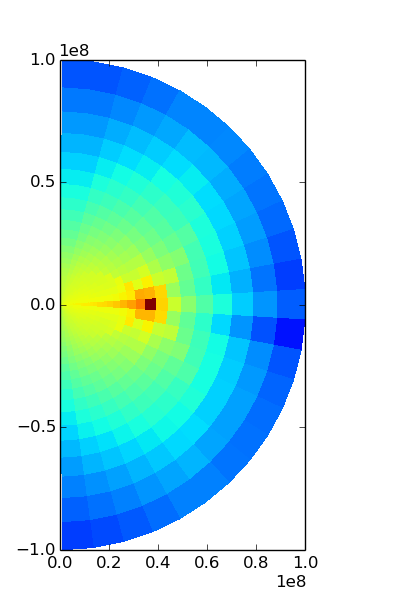}}
\subfigure{\includegraphics[width=0.3\textwidth]{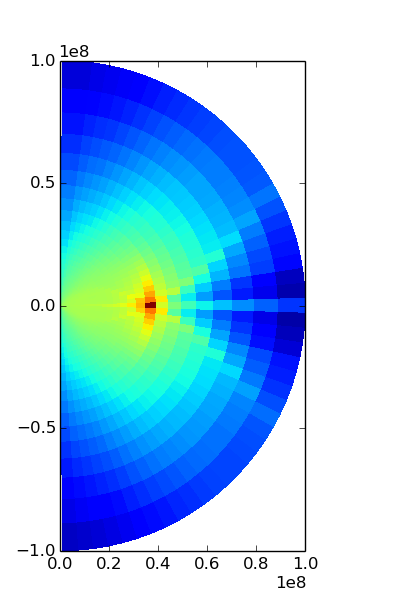}}
\subfigure{\includegraphics[width=0.3\textwidth]{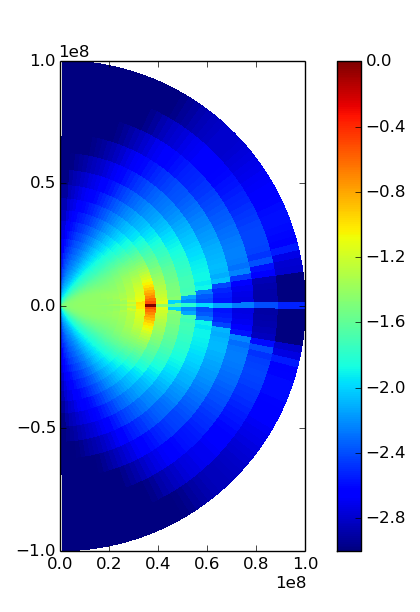}}
\caption{Ray defect pattern from an isotropically emitting ring for 3 different choices of grid resolution.  Here we use a spherical spatial grid with locally defined ray angles.  Panels:  a) $(n_\theta = 26)$ square aspect ratio; b) $(n_\theta = 52)$ 1:2 aspect ratio; c) $(n_\theta = 104)$ 1:4 aspect ratio.  Notice the systematic reduction of radiation near the polar regions when $n_\theta$ increases and also the characteristic ``spider'' pattern arising from our choice of spatial grid interpolation.
\label{fig:spherical-diffusion-compare}}
\end{center}  
\end{figure*}

In Figure \ref{fig:spherical-diffusion-compare}, we show the results
from this diffusion scheme compared to the exact solution for an
off-centre equatorial ring source (e.g. the same setup as described in \S\ref{sec:ringbenchmark}) for different choices of the spatial grid resolution.  We find that our diffusive scheme suffers from systematic overattentuation in the polar regions for spatial grids that too rectangular.  We recommend keeping the grid cell aspect ratio as close to square (1:1) as possible in order to minimize this effect.

The scheme described here is a generalization of that proposed by \citealt{dullemond00}. They applied the same factor $a$ derived above except that they did it solely for the radial ray.  For a central point source, which produces only radial rays, their method and ours produce the same correct inverse-law behaviour of flux. However, when the source is extended, e.g., the spherical shell test problem described in \S\ref{sec:vacuumtest}, their method causes the intensity to fall too steeply. Our method also suffers a similar systematic bias, but it is less severe (see Fig. \ref{fig:vacuumtest}). In the limit of an isotropic radiation field, e.g., the interior of a blackbody enclosure, their method would still cause intensity to decline with increasing radius. Our diffusive method, on the other hand, would give the correct result, viz., no change in intensity. These differences are relatively minor. The main feature of both approaches is that they recover the inverse square law at least approximately.  However, a major limitation of our diffusive approach is that it only works for the case of uniform log-radial grids.  For more complex grids (especially those with nonregular arbitrary/adaptive meshes), the only generalizable solution is to invoke a few iterations of LC to resolve the inverse square falloff from each emitting source.  In particular, if the illumination comes primarily from a very limited set of point sources, then the explicit LC handling of just these point sources shoudn't significantly impact the overall runtime.


\section{Analytic 1D Atmosphere Spectrum}\label{app:analyticSpectrum}

In this problem, we set up a constant flux 1D atmosphere subject to fixed grey opacities ($\kappa_\nu =$ const) with an absorption fraction set to $\epsilon = 10^{-6}$ (i.e. a highly scattering-dominated atmosphere).  The local source function is a combination of thermal emission and reflected light (c.f. Eq \ref{eq:simpleSource}).

The analytic solution to this problem can be easily obtained by considering the moments of the radiative transfer equation.  We define the first few moments of the intensity field as
\begin{align}
J_\nu = \frac{1}{2} \int\limits_{-1}^1 I_\nu d\mu, \nonumber \\
H_\nu = \frac{1}{2} \int\limits_{-1}^1 \mu I_\nu d\mu, \nonumber \\
K_\nu = \frac{1}{2} \int\limits_{-1}^1 \mu^2 I_\nu d\mu, 
\end{align}
where $\mu \equiv \cos(\theta)$ is the angle cosine with respect to the plane normal.  Using these quantities allows us to write the moments of the radiative transfer equation as:
\begin{equation}
\frac{dH_\nu}{d\tau_\nu} = \epsilon(J_\nu-B_\nu) \label{eq:RT1moment}
\end{equation}
\begin{equation}
\frac{dK_\nu}{d\tau_\nu} = H_\nu \label{eq:RT2moment}
\end{equation}
Combining Eqs. \ref{eq:RT1moment} + \ref{eq:RT2moment} and invoking the Eddington approximation ($K_\nu=J_\nu/3$) yields
\begin{equation}
\frac{d^2J_\nu}{d\tau_\nu^2} = 3\epsilon(J_\nu-B_\nu),
\end{equation}
which is simply a constant coefficient second-order inhomogeneous differential equation in $J_\nu$.  This allows us to construct an exact solution using standard methods.  The solutions to the homogeneous system are simply
\begin{align}
\phi_1(\tau) = \exp(\sqrt{3\epsilon}\tau), \nonumber \\
\phi_2(\tau) = \exp(-\sqrt{3\epsilon}\tau).
\end{align}
The particular solution $J_p$ that satisfies the inhomogeneous system is given by\footnote{The particular solution is constructed via the ``variation of parameters'' method}
\begin{equation}
J_p(\tau) = \phi_1(\tau) \int\limits_{\tau}^\infty \frac{\phi_2 g}{W}d\tau' + \phi_2(\tau) \int\limits^{\tau}_0 \frac{\phi_1 g}{W}d\tau',
\end{equation}
where $g = 3\epsilon B$ is the inhomogeneous function, and $W$ denotes the Wronskian, defined by
\begin{align}
W &\equiv \frac{d\phi_1}{d\tau}\phi_2 - \frac{d\phi_2}{d\tau}\phi_1 \\
&= 2\sqrt{3\epsilon}
\end{align}
Putting everything together, the solution takes the form
\begin{equation}
J = J_p + c_1 \phi_1 + c_2 \phi_2,\label{eq:spectralRadSoln}
\end{equation}
where the undetermined constants are set by the boundary conditions of the problem.  At the $\tau \rightarrow \infty$ inner boundary we expect $J \rightarrow B$, so we must eliminate the exponentially growing mode by setting $c_1=0$.  To set the surface boundary condition, we make use of the two-stream approximation and evaluate
\begin{equation}
H(\tau) = \frac{1}{\sqrt{3}}\frac{dJ}{d\tau}
\end{equation}
and enforce a surface boundary condition that is consistent with the Eddington approximation:
\begin{equation}
\frac{H(0)}{J(0)} = \frac{1}{\sqrt{3}},
\end{equation}
which sets 
\begin{equation}
c_2 = - \left(\frac{1 - \sqrt{3}}{1 + \sqrt{3}}\right) \phi_2 \int\limits_0^\infty \frac{\phi_2 g}{W} d\tau'.
\end{equation}
The final step is to determine the thermal source ($B(\tau)$) that is consistent with our radiation solution from Eq. \ref{eq:spectralRadSoln}.  This can be calculated using our two radiative transfer moment equations.  Since we have a constant flux atmosphere, $dH/d\tau = 0$, which implies $J(\tau)=B(\tau)$ from Eq. \ref{eq:RT1moment}.  We combine this with integrating the pressure equation (Eq. \ref{eq:RT2moment}) to yield the full solution 
\begin{align}
J(\tau) &= 3H\tau + J(0) \nonumber \\
\rightarrow B(\tau) &= 3H\left(\tau + \frac{1}{\sqrt{3}}\right) \label{eq:spectralTSoln}
\end{align}
%


\end{document}